# Gas-liquid stratified MHD flows in inclined rectangular ducts

Subham Pal[1] , Ilya Barmak[1,2]*, Arseniy Parfenov[1], Alexander Gelfgat[1], and Neima Brauner[1]

[1]School of Mechanical Engineering, Tel Aviv University, Tel Aviv,6997801, Israel.

[2]Soreq NRC, Yavne, 8180000, Israel.

*Corresponding author. E-mail: ilyab@tauex.tau.ac.il

**Abstract**

This study investigates fully developed stratified gas–liquid magnetohydrodynamic (MHD) flow of an electrically conducting liquid and a nonconducting gas in inclined rectangular ducts subjected to a vertical magnetic field. Analytical and numerical solutions for the velocity and induced magnetic fields are obtained in terms of the governing dimensionless parameters for concurrent upward, concurrent downward, and countercurrent flows. Unlike single-phase MHD flow, duct inclination strongly affects two-phase flow by altering the liquid holdup and the relative contributions of gravitational, frictional, and electromagnetic forces.

The results reveal a complex interplay among gravity, Lorentz forces, and wall and interfacial shear stresses. These interactions govern the liquid holdup, pressure gradient, multiple steady solutions, flooding limits, local backflow, jet-like velocity structures, and pumping requirements. Wall conductivity critically affects the induced magnetic field and Lorentz-force distribution and therefore cannot be neglected, even at very small magnetic Reynolds numbers. Fully insulating ducts generally exhibit the weakest electromagnetic effects and behavior closest to non-MHD flow. Configurations with a conducting bottom wall exhibit substantially stronger electromagnetic effects and greater sensitivity to side-wall conductivity, leading to pronounced changes in the velocity field, liquid holdup, pressure gradient, gas-lubrication effect, and overall pumping-power requirements.

**Keywords:** Magnetohydrodynamics (MHD); gas-liquid; inclined rectangular ducts; walls electrical conductivity; induced magnetic field.

## 1. Introduction

Magnetohydrodynamic (MHD) two-phase duct flows involving a conductive liquid, such as molten metal, electrolyte, or liquid metal alloys, and a non-conductive gas (e.g., air, steam, or inert gases) are encountered in wide range of advanced engineering and industrial systems. The motion of conductive liquid is affected by an external electromagnetic field, while the gas phase is not, offering useful ways to control flow, heat transfer, and phase distribution. Applications involving gas-liquid MHD flows are found in energy and power systems, metallurgy and material processing, aerospace and space application, chemical and processing systems, cooling systems, microfluidics and lab-on-chip devices (e.g., [1-8]) For example, since molten metals (like sodium, lithium, aluminum, magnesium) are highly reactive, a flow of non-conductive inert gas (e.g., argon) is introduced to maintain gas blanket above liquid metal protecting it chemically and thermally by preventing oxidation, contamination, and unwanted chemical reactions [9-10]. In other applications, the gas is injected to form a dispersed phase. However, due to the high density difference between the liquid metal and gas, they may separate to form stratified flow. Even though the gas is non-conductive, and the electric current and MHD interactions occur only in the liquid metal, the gas affects the flow of the conductive liquid through shear at the gas-liquid interface and pressure coupling. These in turn determine the two-phase flow configuration and thereby the MHD flow characteristics.

Most research on MHD flows has focused on single-phase flows in rectangular ducts, a common geometry in practical applications. Early works considered the two-plate (TP) geometry, with the seminal study of Hartmann [11] who demonstrated that the axial velocity profile depends solely on the Hartmann number, $Ha = B_0 H \sqrt{(\sigma/\eta)}$, where $H$ is the channel height, $B_0$ is the external transverse magnetic field strength, and $\eta, \sigma$ are the fluid viscosity and electric conductivity, respectively. However, Hartmann's solution neglects the magnetic field induced by the fluid motion. As an electrically conducting fluid moves through a magnetic field, electric currents are generated perpendicular to the field. These currents, in turn, produce an induced magnetic field and a corresponding axial Lorentz force that affect the pressure gradient required to sustain the flow. For the case of channel with electrically insulating walls, Shercliff [12] derived a solution that accounts for the induced magnetic field. On the other hand, Hartmann's solution, in fact, applies to the limiting case of perfectly conducting walls. It has been shown (e.g., [4]) that in both cases the normalized velocity profile depends only on the Hartmann number and becomes identical when scaled by the corresponding maximum velocity. Nevertheless, for a given Hartmann number and imposed pressure gradient, the volumetric flow rate and maximum velocity are greater in channels with electrically insulating walls than in channels with perfectly conducting walls.

Because the electrical conductivity of the duct walls is an important design parameter in many MHD applications, its effect on flow characteristics has been extensively investigated. Hunt [13] derived analytical solutions for the two-dimensional (2D) velocity field in rectangular ducts, accounting for the influence of wall electrical conductivity on both the velocity and induced magnetic fields. Numerous other works have subsequently investigated analytical solutions for fully developed, laminar, single-phase MHD flows in channels and ducts subjected to transverse magnetic fields (e.g.,[14-17]). These studies have been reviewed in several articles and book chapters (e.g., [4], [18-20]).

Most analytical solutions are expressed as infinite series of hyperbolic functions. Although exact, these solutions become increasingly difficult to evaluate as the Hartmann number increases, and asymptotic approximations are therefore commonly employed in the high-Hartmann-number regime (e.g., [4], [21]). Using numerical solutions of the governing equations, Krasnov et al. [22] and Boeck and Krasnov [23] demonstrated that, for electrically insulating rectangular ducts, the velocity profiles, normalized by their respective maximum velocities, closely resemble those predicted by the two-plate model at the same Hartmann number. However, accurate numerical simulations require very fine spatial resolution near the duct walls in order to resolve the steep velocity gradients within the thin Hartmann (perpendicular to the external magnetic field) and Shercliff (parallel to the external field) boundary layers and to accurately predict the associated wall shear stresses and the Lorentz force.

The presence of a gas layer in gas–liquid MHD flows can significantly alter the flow characteristics of the conductive liquid, as well as the pressure gradient and pumping-power requirements. However, considerably fewer studies have addressed two-phase MHD flows, and most of them have been limited to the simple two-plate (TP) geometry (e.g., [24-27]). In all of these studies, the flow configuration was prescribed a priori, with both the in-situ holdup of the heavier liquid layer and the pressure gradient specified as input parameters, even though these quantities are generally unknown and should be determined as part of the solution. More recently, Parfenov et al. [28] presented a model for predicting the in-situ holdup and pressure gradient in horizontal and inclined channels for specified flow rates of a conducting liquid and a gas phase. Their solution enabled assessment of the potential reduction in pressure gradient and the associated pumping-power savings resulting from gas injection. However, none of the aforementioned studies considered the effect of channel-wall electrical conductivity on the induced magnetic field and its consequent influence on the velocity profile of the conductive liquid, the pressure gradient, and the overall two-phase flow characteristics.

In our recent studies of gas–liquid MHD flow, we investigated the influence of duct-wall electrical conductivity on the induced magnetic field and the resulting flow characteristics in the two-plate (TP) geometry and in rectangular ducts [29],[30]. The presence of a non-conductive layer (e.g., gas) breaks the symmetry of the boundary conditions (cf. single-phase MHD flows), thereby affecting not only the velocity of the conductive liquid, but also the induced magnetic field and the Lorentz force acting on the flow. Consequently, the dependence of the liquid-flow characteristics and the associated pressure gradient on the duct aspect ratio, wall-conductivity configuration (i.e., the electrical conductivities of the bottom and side walls), and on the strength and orientation of the applied magnetic field differs significantly from that observed in single-phase flow of the same conductive liquid at the same flow rate and in the same duct geometry.

The solutions developed by Barmak et al. [29] and Pal et al. [30] were employed to quantify the influence of the gas phase on the in-situ liquid holdup, velocity field, pressure gradient, flow-lubrication effects, and pumping-power requirements. The results demonstrated that, regardless of the magnetic Reynolds number, the flow characteristics are strongly influenced by both the wall-conductivity configuration and the orientation of the external magnetic field. However, these studies were limited to horizontal gas–liquid systems, for which the flow characteristics are independent of the liquid-to-gas density ratio. In practical applications, even systems designed to operate under horizontal-flow conditions may exhibit slight inclinations, potentially introducing significant gravitational effects and modifying the flow behavior. While the gravity force does not affect the flow characteristics in single-phase flow in inclined closed ducts, in free surface flow it determines the thickness of the conductive liquid (e.g., [31], [32]).

Stratified gas–liquid flows are gravity-dominated and highly sensitive to duct inclination. The pronounced effect of even small inclination from the horizontal on the two-phase flow configuration, pressure gradient, and other flow characteristics in the absence of a magnetic field, is well documented in the two-phase flow literature (e.g., [33], [34]) and it becomes more pronounced as the liquid-to-gas density ratio increases. The role of gravity also differs fundamentally between concurrent downward, concurrent upward, and countercurrent flows. The possibility of gravity-driven multiple solutions in inclined channels for a given two-phase system and fixed operating conditions has also been established in the literature [35-37] and may lead to operational instabilities [34]. In countercurrent flows, two distinct solutions for the liquid holdup are always possible, whereas in concurrent upward- and downward-inclined flows, up to three solutions may exist within a limited range of flow parameters (e.g., [38-40]). However, to the best of our knowledge, the flow characteristics of stratified gas–liquid MHD flow in inclined rectangular ducts have not yet been investigated. In particular, the combined influence of duct-wall electrical conductivity and the associated Lorentz force in such gravity-dominated configurations remains unclear.

In the present study, we consider gas–liquid stratified MHD flows in inclined rectangular ducts. Numerical solutions of the 2D governing equations for steady, fully developed flows are obtained for the velocity and induced magnetic field distributions under various combinations of bottom and side-wall electrical conductivities. Analytical solutions are derived for selected wall-conductivity configurations, as well as for the simplified TP model, and are used to verify the corresponding numerical results. The verified numerical framework is then employed to investigate the combined effects of shallow duct inclination, wall-conductivity configuration on the resulting flow characteristics in both concurrent upward and concurrent downward inclined flows.

## 2. Problem Formulation

The configuration of stratified MHD flow of a non-conductive gas and conductive liquid metal in an inclined rectangular duct of the high $H$ and width $W$ (i.e., aspect ratio $A = W/H$) is shown in Fig. 1. The inclination angle is denoted as $\beta$ $(0 < \beta < \pi/2)$. The flow is driven by a pressure gradient imposed in the axial direction of the duct ($z$-direction) and a gravity force acting downwards. The MHD flow is also subject to a constant external magnetic field, $B_0$, acting in the $y$-direction, i.e., perpendicular to the bottom wall. The superficial velocities of the heavy (liquid metal) and light (gas) phases are $U_{1S}$ and $U_{2S}$, respectively, where $U_{1S,2S} = Q_{1,2}/(HW)$ and $Q_{1,2}$ are the corresponding flow rates ($Q_{21} = Q_2/Q_1$). In this work, the flow is assumed to be steady fully developed laminar and unidirectional in the $z$-direction, so that the dimensionless velocity is defined as $\boldsymbol{U} = (0,0,U(x,y))$ and all the flow and electromagnetic variables are independent of $z$. There are three distinct flow configurations that can be classified according to the signs of $U_{1S}$ and $U_{2S}$: downward inclined flow – both superficial velocities are positive, countercurrent flow where $U_{1S} > 0$ and $U_{2S} < 0$, and upward inclined flow where both superficial velocities are negative.

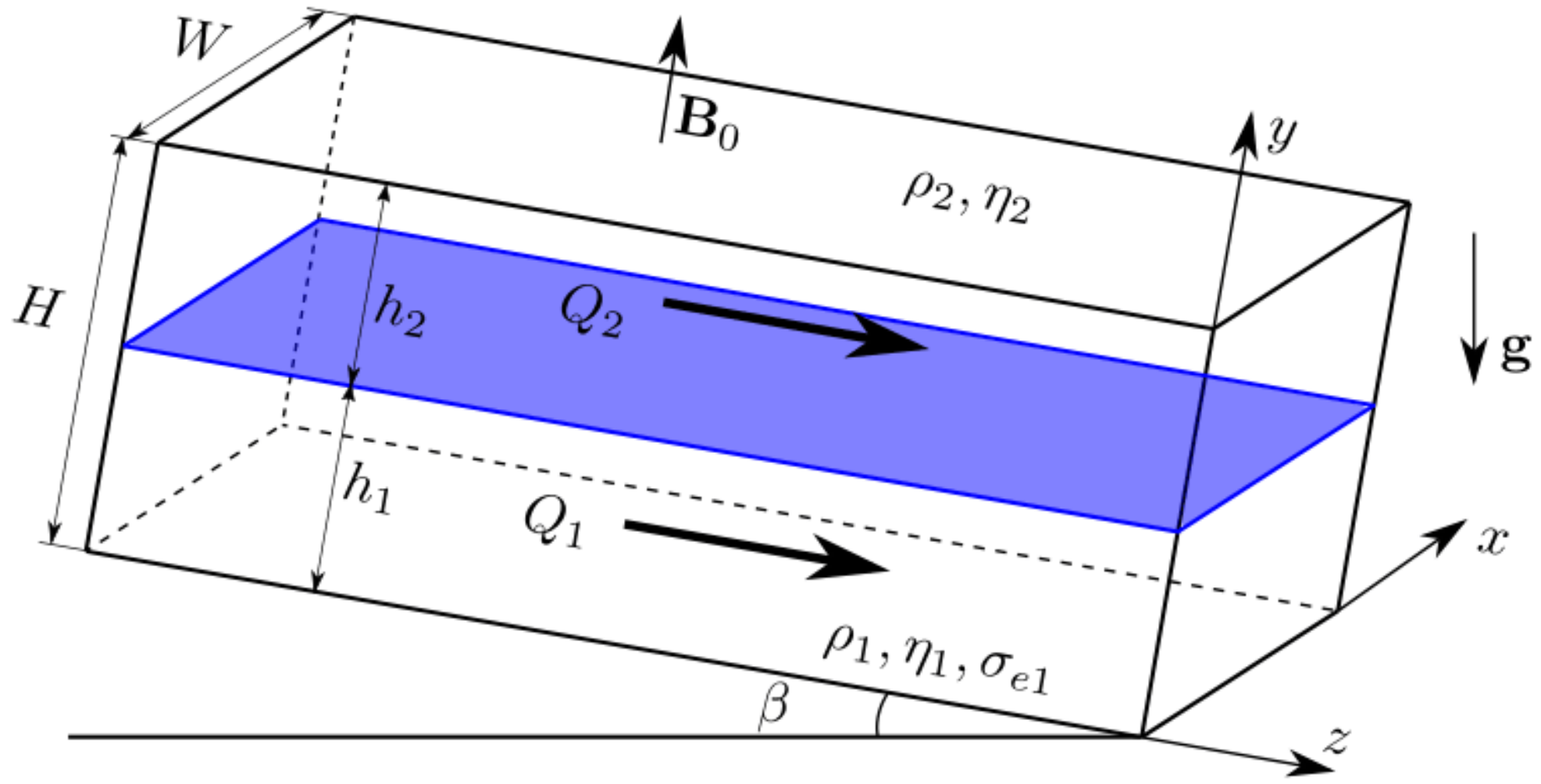

*Figure 1. Schematics of stratified two-phase MHD flow in an inclined rectangular duct.*

The gas-liquid interface is assumed to be plane and the height of the lower phase layer is denoted as $h_1$, while the holdup is defined as $h = h_1/H$. The physical properties of the conductive liquid include the dynamic viscosity ($\eta_1$), its density ($\rho_1$) and its electric conductivity ($\sigma_{e1}$). The dimensionless form of the only non-zero axial components of the momentum and induction equations for the conductive phase and the momentum equation for the gas phase then read:

$$\frac{\partial^2 U_1}{\partial x^2} + \frac{\partial^2 U_1}{\partial y^2} + Ha\,\frac{\partial b}{\partial y} = \tilde{G}_1, \qquad (1)$$

$$\frac{\partial^2 b}{\partial x^2} + \frac{\partial^2 b}{\partial y^2} + Ha\,\frac{\partial U_1}{\partial y} = 0, \tag{2}$$

$$\frac{\partial^2 U_2}{\partial x^2} + \frac{\partial^2 U_2}{\partial y^2} = \eta_{12}\tilde{G}_2. \tag{3}$$

In these equations, the lengths are scaled by $H$, the velocity is scaled by the superficial velocity of the conductive liquid, $U_{1S}$. The driving force in inclined ducts includes both the imposed pressure gradient and a phase-dependent gravitational force, $\tilde{G}_{1,2} = \frac{H^2}{\eta_1 U_{1s}}\left(\frac{dp}{dz} - \rho_{1,2} g sin\beta\right)$, and $\eta_{12}$ is the liquid-to-gas viscosity ratio. The difference between $\tilde{G}_2$ and $\tilde{G}_1$ is equal to the system inclination parameter, $\tilde{Y}$:

$$\tilde{Y} = \tilde{G}_2 - \tilde{G}_1 = \frac{(\rho_1 - \rho_2) g sin\beta H^2}{\eta_1 U_{1S}}. \tag{4}$$

The induced magnetic field, $b$, is scaled by $B_0 Re_m/Ha$, with $Re_m = \mu_0 \sigma_{e1} U_{1S} H$ and $Ha = B_0 H\sqrt{\sigma_{e1}/\eta_1}$ as magnetic Reynolds and Hartmann numbers, respectively. Note that the sign of $Re_m$ is negative in case of upward flow of the conductive liquid ($U_{1s} < 0$). Accordingly, in all flow configurations the induced magnetic field $b$ is positive when it is acting in the direction of $U_{1s}$. Therefore, $\frac{\partial b}{\partial y} > 0$, indicates that the Lorenz force is acting in the direction of the conductive liquid flow.

The velocity in Eqs.(1)-**Error! Reference source not found.** is subject to no-slip conditions on the duct walls and continuous across the fluid-fluid interface, where continuity of the shear stress takes place as well:

$$U_1(x; y = 0) = 0,\;\; U_1(x = 0, AR; y) = 0, \tag{5}$$

$$U_2(x; y = 0) = 0,\;\; U_2(x = 0, AR; y) = 0, \tag{6}$$

$$U_1(x; y = h) = U_2(x; y = h), \qquad \eta_{12}\left.\frac{\partial U_1}{\partial y}\right|_{x;y=h} = \left.\frac{\partial U_2}{\partial y}\right|_{x;y=h}. \tag{7}$$

Both the velocity and shear stress vary along the interface (i.e., in the lateral $x$-direction), due to the presence of side walls. They also affect the induced magnetic field, which is subject to the following boundary on the duct walls and the interface (see details in [29],[30]):

$$-\left.\frac{\partial b}{\partial y}\right|_{x;y=0} + \frac{1}{c_{bw}} b(x; y = 0) = 0, \tag{8}$$

$$-\left.\frac{\partial b}{\partial x}\right|_{x=0,AR;y} + \frac{1}{c_{sw}} b(x = 0, AR; y) = 0, \tag{9}$$

$$b(x; y = h) = 0. \tag{10}$$

where $c_{bw}$ and $c_{sw}$ represent the electric conductivities of the bottom and side walls, respectively. In the limiting case of a fully insulating bottom (side) wall $c_{bw} = 0$ ($c_{sw} = 0$), whereas in the limiting case of perfectly conducting bottom (side) wall $c_{bw} \to \infty$ ($c_{sw} \to \infty$). In the former case, the boundary condition reduces to $b = 0$, whereas in the latter it reduces to the zero normal derivative of $b$ at the corresponding wall. Accordingly, at the non-conducting gas-liquid interface $b = 0$, and the conductivity of the top does not affect the gas-liquid flow characteristics.

The total pressure gradient, $dp/dz$ can be represented by the sum of the frictional and gravitational pressure gradients. In the dimensionless form:

$$\tilde{P} = \frac{dp/dz}{-\eta_1 U_{1S}/H^2} = \frac{1}{-\eta_1 U_{1S}/H^2}\left[\frac{dp_f}{dz} + \frac{dp_g}{dz}\right] = \tilde{P}_f + \tilde{P}_g \tag{11}$$

$$\tilde{P}_f = -\frac{\frac{dp_f}{dz}}{\eta_1 U_{1S}/H^2} \tag{11.1}$$

$$\tilde{P}_g = -\frac{1}{\eta_1 U_{1S}/H^2}\left[\{\rho_1 h + \rho_2(1-h)\}gsin\beta\right] = -\left(h + \frac{\rho_2}{\rho_1 - \rho_2}\right)\tilde{Y} \tag{11.2}$$

Note that in Eq. (11), a minus sign was introduced in the normalization of the pressure gradient to reflect that the frictional pressure gradient is opposite to the flow direction. The dimensionless driving force in each phase is then:

$$\tilde{G}_1 = \frac{H^2}{\eta_1 U_{1s}}\left[\frac{dp_f}{dz} - (1-h)(\rho_1 - \rho_2)gsin\beta\right] = \left[-\tilde{P}_f - (1-h)\tilde{Y}\right] \tag{12.1}$$

$$\tilde{G}_2 = \frac{H^2}{\eta_1 U_{1s}}\left[\frac{dp_f}{dz} + h(\rho_1 - \rho_2)gsin\beta\right] = \left[-\tilde{P}_f + h\tilde{Y}\right] \tag{12.2}$$

Whereby:

$$\tilde{P}_f = -\tilde{G}_1 - (1-h)\tilde{Y} = -\tilde{G}_2 + h\tilde{Y};\ \tilde{P}_g = -\left[h + \frac{\rho_2}{\rho_1 - \rho_2}\right]\tilde{Y} \tag{13}$$

A gas lubrication factor, $\tilde{P}_f^1$ is introduced to evaluate the potential of reducing the pressure gradient required to transport the conductive liquid by introducing a gas flow. It is obtained by scaling $\tilde{P}_f$ by the frictional pressure gradient obtained in a single-phase flow of the conductive liquid for the same $Ha$, $\tilde{P}_{f_{1s}}$:

$$\tilde{P}_f^1 = \frac{(dp_f/dz)}{(dp_f/dz)_{1s}} = \frac{\tilde{P}_f}{\tilde{P}_{f_{1s}}} \tag{14}$$

A pressure gradient reduction factor which represents the effects of both the inclination and the gas flow on the pressure gradient and pumping power is $\tilde{P}^1 = \tilde{P}/\tilde{P}_{f_{1s}}$. To examine the effect of the gas flow only, the total pressure gradient should be compared to the total pressure gradient in single-phase flow with the same $Ha$ and $\beta$, $(dp/dz)_{1s}$:

$$\left(\frac{dp}{dz}\right)_{1S} = \left(\frac{dp_f}{dz}\right)_{1s} + \rho_1 gsin\beta \tag{15}$$

thereby the potential for total pressure and power reduction due to the gas flow are presented by the following factors:

$$\tilde{P}^{1T} = \tilde{P}^{1}\left[1-\frac{\rho_1}{\rho_1-\rho_2}\frac{\tilde{Y},}{\tilde{P}_{f_{1S}}}\right]^{-1} \qquad (16)$$

$$Po^{T} = \tilde{P}^{1T}\frac{U_{1S}+U_{2S}}{U_{1S}} = \tilde{P}^{1T}(1+Q_{21}) \qquad (17)$$

## 3. Numerical solution

In order to solve the coupled governing equations (1)-**Error! Reference source not found.** numerically for given superficial velocities $U_{1S}$ and $U_{2S}$ and to find the required pressure gradient and conductive liquid holdup, the velocity field in each phase, $U_1(x,y)$ and $U_2(x,y)$, as well as the induced magnetic field, $b(x,y)$, can be conveniently split into the superposition of the component flow driven solely by the imposed pressure gradient $\tilde{G} = \frac{dp/dz}{\eta_1 U_{1S}/H^2}$ (denoted by superscript $^{\pi}$) and the other one that driven by gravity only (denoted by superscript $^{\gamma}$):

$$U_{1,2} = \tilde{G}U_{1,2}^{p} - U_{1,2}^{g} \qquad (18)$$

$$b = \tilde{G}b^{p} - b^{g} \qquad (19)$$

where $U_{1,2}^{p}$, $U_{1,2}^{g}$, $b^{p}$, and $b^{g}$ are solutions of the following equations:

$$\frac{\partial^2 U_1^p}{\partial x^2}+\frac{\partial^2 U_1^p}{\partial y^2}+Ha\,\frac{\partial b^p}{\partial y}=1 \qquad (20.1)$$

$$\frac{\partial^2 b^p}{\partial x^2}+\frac{\partial^2 b^p}{\partial y^2}+Ha\,\frac{\partial U_1^p}{\partial y}=0 \qquad (20.2)$$

$$\frac{\partial^2 U_2^p}{\partial x^2}+\frac{\partial^2 U_2^p}{\partial y^2}=\eta_{12} \qquad (20.3)$$

$$\frac{\partial^2 U_1^g}{\partial x^2}+\frac{\partial^2 U_1^g}{\partial y^2}+Ha\,\frac{\partial b^g}{\partial y}=\frac{Re}{Fr^2}\sin\beta=\frac{\rho_{12}}{\rho_{12}-1}\tilde{Y} \qquad (21.1)$$

$$\frac{\partial^2 b^g}{\partial x^2}+\frac{\partial^2 b^g}{\partial y^2}+Ha\,\frac{\partial U_1^g}{\partial y}=0 \qquad (21.2)$$

$$\frac{\partial^2 U_2^g}{\partial x^2}+\frac{\partial^2 U_2^g}{\partial y^2}=\frac{\eta_{12}}{\rho_{12}}\frac{Re}{Fr^2}\sin\beta=\frac{\eta_{12}}{\rho_{12}-1}\tilde{Y} \qquad (21.3)$$

The boundary conditions for these variables are the same as for the total velocity and induced magnetic fields and specified above in Eqs. (5)-(7), (8)- (8).

Following our previous works [29],[30], we adopt the finite volume method with the duct cross section divided into $N_x$ and $N_y$ quadrilateral cells in $x$- and $y$-directions, respectively, and using the sine-function stretching near the wall and the interface (in y-direction). Thereby the cell center $y$ coordinates are obtained from originally uniformly distributed grid cells between 0 and 1 for each sublayer (phase) as $y \to y - a_y \sin(2\pi y)$. In this work, we use $a_y = 0.06$ and the same stretching in the $x$-direction. The resulting sparse linear equation systems obtained after discretization are solved by a direct solver from MUMPS package. More details on the numerical method and its verification for horizontal MHD flows can be found in [29],[30].

However, instead of 3 governing equations (Eqs. (1)-(3)), which were solved in horizontal flow, in inclined flows the system of 6 coupled governing equations (20)-(21) with the corresponding boundary conditions is solved with varying holdup until its proper value, along with pressure gradient $\tilde{G}$, is found by the secant method to satisfy the prescribed values of superficial velocities $U_{1S,2S}$:

$$U_{1S} = \tilde{G} U_{1S}^{p} - U_{1S}^{g} \qquad (22.1)$$

Hence:

$$\tilde{G} = \frac{U_{1S} + U_{1S}^{g}}{U_{1S}^{p}} \qquad (22.2)$$

and

$$U_{2S} = \tilde{G} U_{2S}^{p} - U_{2S}^{g} \qquad (22.3)$$

where

$$U_{1S}^{0,g} = \int_0^h \int_0^{AR} U_1^{p,g} dxdy, \quad U_{2S}^{0,g} = \int_h^1 \int_0^{AR} U_2^{p,g} dxdy. \qquad (23)$$

Once the holdup and pressure gradient are found, the total velocity field and induced magnetic field are calculated by Eqs.**Error! Reference source not found.** and **Error! Reference source not found.**, respectively.

To verify the numerical method that allows one to obtain solutions for the general case of wall conductivities and inclined flow configurations, the analytical solution is used which could be derived for rectangular ducts with insulating sidewalls and arbitrary bottom wall conductivity (see Appendix A). Calculation of the analytical solution involves evaluation of series of hyperbolic functions. Performing these calculations with sufficient precision requires taking into account up to 200 terms, when each term (hyperbolic function) is evaluated with double floating-point precision (i.e., 15-16 decimal places).

The convergence of the numerical solution with increasing number of computational cells, and comparison with the analytical solution results is reported in Table 1. In this exercise we compare the flow rate ratio calculated for a given holdup in inclined square ducts with different wall conductivities. It is seen that the numerical method exhibits a convergence to within the third decimal place. The numerical results coincide with the analytical solution also within the three decimal places, which is reasonable for the lower order finite volume method applied.

*Table 1. Grid convergence for the numerical solution for MHD mercury-air flows in inclined square ducts at Ha=103.625 with different wall conductivity configurations.*

| $N_x$ | $N_y$ | $Q_{21}$ | | | |
|---|---|---|---|---|---|
| | | $U_{1S} = -0.005, h = 0.9, \beta = 0.2$ | | $U_{1S} = 0.1, h = 0.9, \beta = 5$ | |
| | | Concurrent upward flow | | Counter-current flow | |
| | | IbIs | CbIs | IbIs | CbIs |
| 400 | 400 | 153.1348 | 161.6933 | -188.9344 | -175.0005 |
| 400 | 600 | 153.3508 | 162.0238 | -189.2004 | -175.3501 |
| 400 | 800 | 153.4614 | 162.1920 | -189.3365 | -175.5282 |
| 800 | 800 | 153.4589 | 162.1896 | -189.3335 | -175.5247 |
| 400 | 1000 | 153.5285 | 162.2938 | -189.4193 | -175.6360 |
| 1000 | 1000 | 153.5257 | 162.2912 | -189.4158 | -175.6322 |
| **Analytical** | | **153.6445** | **162.5518** | **-189.560** | **-175.8791** |

## 4. Results and Discussion

The holdup and dimensionless frictional pressure gradient ($\tilde{P}_f$) in inclined rectangular ducts are determined by a set of five dimensionless parameters ($\eta_{12}, Q_{21}, Ha, \tilde{Y}, AR$), as well as by the wall conductivities. Calculation of the total (dimensionless) pressure gradient factor $\tilde{P}$ (see Eq. **Error! Reference source not found.**) requires taking into account the hydrostatic pressure gradient contribution, thereby introducing dependency on the density ratio, $\rho_{12} = \rho_1/\rho_2$. Moreover, depending on the values of $Q_{21}$ and $\tilde{Y}$, the liquid metal and gas can form concurrent downward, concurrent upward, or countercurrent flow configurations.

*Table 2. Mercury and air physical properties*

| Mercury | density | $\rho_1 = 1.35 \cdot 10^4$ kg/m$^3$ |
|---|---|---|
| | dynamic viscosity | $\eta_1 = 1.49 \cdot 10^{-3}$ Pa s |
| | electric conductivity | $\sigma_{e1} = 10^6\ 1/\Omega \cdot$ m |
| Air | density | $\rho_2 = 1$ kg/m$^3$ |
| | dynamic viscosity | $\eta_2 = 1.8 \cdot 10^{-5}$ Pa s |
| | electric conductivity | $\sigma_{e2} = 0$ |

To enable a meaningful parametric study given the large number of governing parameters and three possible inclined flow configurations, a mercury-air system in a square duct was selected to represent a typical two-phase MHD flow. The relevant physical properties of the fluids are listed in Table 2. Thus, the following values are fixed: $\eta_{12}$ =(81.87) and $\rho_{12}$(=13534) and $AR$=1. The value of the magnetic Prandtl number of mercury, $Pr_m = Re_m/Re_1 = (\mu_0\sigma_{e1})/(\rho_1/\eta_1)$, is $1.38735 \cdot 10^{-7}$. Considering the relevant practical range of bulk Reynolds number, $Re_1 = \rho_1 U_{1s} H/\eta_1$, the magnetic Reynolds number of the considered flow corresponds to $Re_m \ll 1$ . However, as already demonstrated by the two-plate analytical solution, the induced magnetic field affects the flow characteristics even for $Re_m \ll 1$. Therefore, its dependence on the walls' conductivity should be considered.

We present results for the two limiting cases of perfectly insulating and ideally conducting (bottom or side) walls, as these provide physically meaningful lower and upper bounds on the influence of wall conductivity. Together, these limits delineate the range of expected variation of the flow characteristics with the change of the walls' conductivities. The considered range of the intensity of external magnetic field is $B_0 \in [0,0.2]$Tesla, which represents standard laboratory conditions, whereby the corresponding Hartmann numbers are $Ha \in [0,103.62]$ in a H=0.02m duct.

The objective of the present study is to examine the influence of duct inclination on gas-liquid MHD flows. While the duct inclination does not influence single-phase MHD flows, it significantly affects two-phase flows by altering the phase distribution and the overall flow behavior. As a reference for understanding these effects, Figure 2 demonstrates the influence of the magnetic field and wall conductivities on the variation of holdup and pressure gradient changes with $Q_{21}$ in horizontal square duct, $\tilde{Y} = 0$ (Pal et al., 2026). Given the bottom and side walls' conductivities and the Hartmann number, the holdup is uniquely determined by the flow rate ratio.

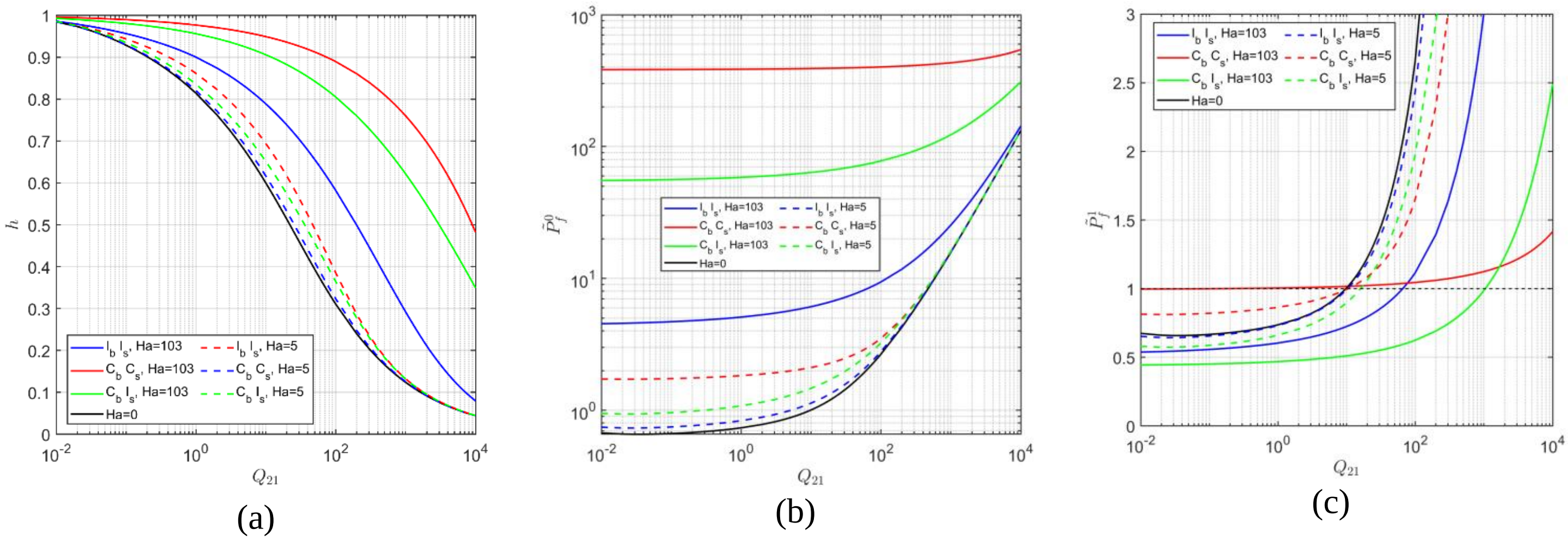

**Figure 2** Horizontal square duct (a) holdup (b) pressure gradient, $\tilde{P}_f^0 = \tilde{P}_f/28.455$ , (c) Lubrication factor $\tilde{P}_f^1$.

Figure 2a clearly shows that applying a magnetic field (i.e., non-zero $Ha$) increases the conductive liquid holdup (Figure 2a). This is attributed to the retarding Lorentz force, which slows down the liquid. Consequently, a larger flow cross section occupied by the conductive liquid is required to maintain a prescribed flow rate ratio compared to the case of $Ha = 0$ (i.e., no external magnetic field). However, this figure also demonstrates the pronounced influence of the wall conductivities on both the holdup and pressure gradient. Obviously, since only the lower phase is conductive, the conductivity of the upper wall has of no effect. The results indicate that for the same $Q_{21}$ the holdup and pressure gradient (Figure 2b) are significantly lower in a fully insulated (IbIs) duct than in a perfectly conducting duct (CbCs). As a result, a much higher $Q_{21}$ (i.e., higher gas flow rate) and larger pressure gradient are required to maintain the same holdup. Note that the $\tilde{P}_f^0$ (shown in Figure 2b) represents the frictional pressure gradient normalized by the that obtained in single-phase flow in a square duct for the same liquid flow rate $\left(dp_f/dz\right)_{1s} = -28.455\eta_1 U_{1S}/H^2$.

The lubrication factor, defined as the ratio between the axial pressure gradient in two-phase flow and that in single-phase flow of the conductive fluid in the same duct for the same *Ha* and wall configuration conductivity, is shown in Figure 2c. The $Q_{21}$range where $\tilde{P}_f^1 < 1$ corresponds to conditions where the pressure drop can be reduced by introducing an air flow into the duct. In a fully conducting square duct, the lubrication effect is weak and limited to low *Ha*. On the other hand, in a perfectly insulated (IbIs) duct, the air lubrication becomes stronger with increasing *Ha*, and at high *Ha*, air lubrication is more pronounced and extends over a wider range of flow rates than in the absence of magnetic field (*Ha*=0).

Figure 2(a-c) further illustrates a negligible effect of the side wall conductivity on holdup when the bottom wall is insulating. However, their effect becomes significant when the bottom wall is conducting. In fact, the strongest lubrication effect, characterized by the lowest $\tilde{P}_f^1$ and over the widest range of $Q_{21}$, is obtained in the CbIs, where the bottom wall is conductive, but the side walls are insulating. The influence of the wall-conductivity configuration on the two-phase MHD flow characteristics arises from differences in the electric currents, the resulting induced magnetic field, and the associated Lorentz force acting on the conductive liquid under the various wall-conductivity configurations. Importantly, these effects remain independent of $Re_m$ (see Pal et al., 2026).

The analytical solutions derived for the TP model provide valuable insights into the influence of the system parameters on the flow characteristics. However, the resulting values should not be considered as quantitatively reliable predictions for practical ducts of finite aspect ratio, typically around $AR = 1$ (square duct). This limitation is illustrated in Figure 3, which shows the flow rate ratio and pressure gradient values obtained for a square duct to the corresponding values predicted by the TP model for the same holdup and channel inclination. Since the TP model ignores the side-wall effects, these characteristics are shown only for perfectly conducting or fully insulating ducts.

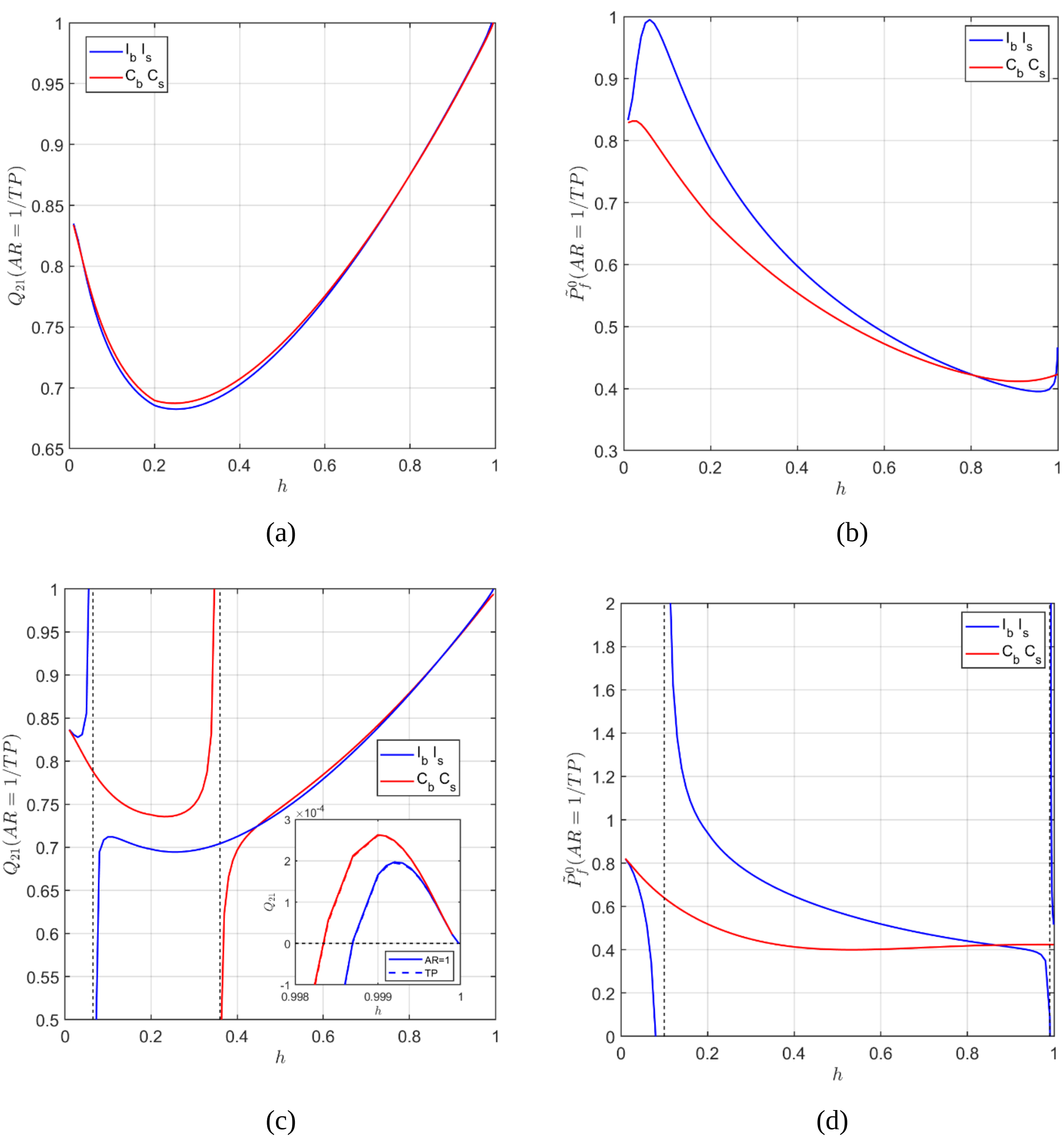

**Figure 3** Ratio of $Q_{21}$ and frictional pressure gradient values vs. the holdup obtained for a insulating square duct to the corresponding values predicted by the TP model for insulating bottom wall, *Ha*=103.625 (a,b) Concurrent up flow, ($\beta$ =1°, $U_{1s} = -0.005 m/s$ , $\tilde{Y} = -124025$) (c,d) Concurrent down and countercurrent flow ($\beta$ =5°, $U_{1s} = 0.1$ *m/s*, $\tilde{Y} = 30968.5$)

For upward shallow inclination ($\tilde{Y} < 0$, Figure 3(a,b), the TP model is seen to substantially overestimate $Q_{21}$, while underestimating the pressure gradient, particularly at relatively low holdups ($h$<0.6). Similar trends are observed for both conducting and insulating ducts. For $\tilde{Y} > 0$ (Figure 3c,d), the variation of the TP-model error with the holdup is more complex. In this case, a major part of the holdup range corresponds to countercurrent flow ($Q_{21}$<0), whereas the concurrent downward flow ($Q_{21}$>0) is restricted to the low holdup range, except for high holdups (close to 1) where additional two solutions may be obtained (Parfenov et al. 2024). In general, the TP-model holdup over predicts the holdup also for $\tilde{Y} > 0$ (Figure 3c). However, the transition to counter current flow ($Q_{21}$=0) does not correspond to the same holdup in the TP model and in square duct solution. Consequently, the variation of the relative error in $Q_{21}$ exhibits discontinuity at the holdup corresponding $(Q_{21})_{TP}$ =0 leading to very large errors in its vicinity. When triple solutions exist in the concurrent downward flow, and additional $Q_{21}$=0 crossing occurs also at high holdups. In the case shown in Figure 3c, however, this crossing occurs at practically the same location in the square duct and in the TP-model solutions and does not involve appreciable errors in the TP model predictions (see insert of Figure 3c, where the TP and the AR=1 curves practically coincide). Figure 3d shows that for h<0.6, the TP-model substantially underestimates the frictional pressure gradient in a conducting duct. In insulating duct, the $\tilde{P}_f$ is also underestimated by the TP model over most of the countercurrent flow region. However, in this case $\tilde{P}_f$=0 at two points along the holdup curve: one at low holdup corresponding to countercurrent flow and the other one for high holdups (close to 1) corresponding to concurrent downward flow (see section 5.2 below). The TP model also predicts these zero crossing, but not at the same holdup values. As a result, discontinuity in the $\tilde{P}_f$ ratio occurs where the TP model predicts $\tilde{P}_f$=0 and producing large positive and negative errors in their vicinity.

### 4.1. Concurrent upward flow in square ducts

Concurrent upward flow corresponds to negative values of the superficial velocities $U_{1S,2S} < 0$ (see Figure 1), hence $Q_{21} > 0$ and $\tilde{Y} < 0.$ Due to the large density difference between liquid metals and the gas phase, the flow characteristics are expected to be highly sensitive already for shallow upward inclinations.

Figure 4 illustrates the effect of the external magnetic field intensity (i.e., $Ha$) and walls' conductivities on the variation of the holdup with the air-to-mercury flow-rate ratio in a square duct (H=0.02m) for low mercury flow rate $U_{1s} = -0.005$ m/s, and $\beta$ =0.2°, 0.5°, 1°. These correspond to negative inclination parameters, such that $-\tilde{Y} = 24806$, 62015, 124025, respectively. The obtained results are dimensionless and valid for all $U_{1s}$ and $\beta$ that result in the same value of $\tilde{Y}$. A comparison between results obtained for $Ha = 0$ and $Ha = 103.625$ delineates the effects of the magnetic field intensity and sensitivity to the wall conductivities in the range of to $B_{0y}$ =0.2 Tesla. The variation of $Q_{21}$ corresponds to the change in the air flow rate.

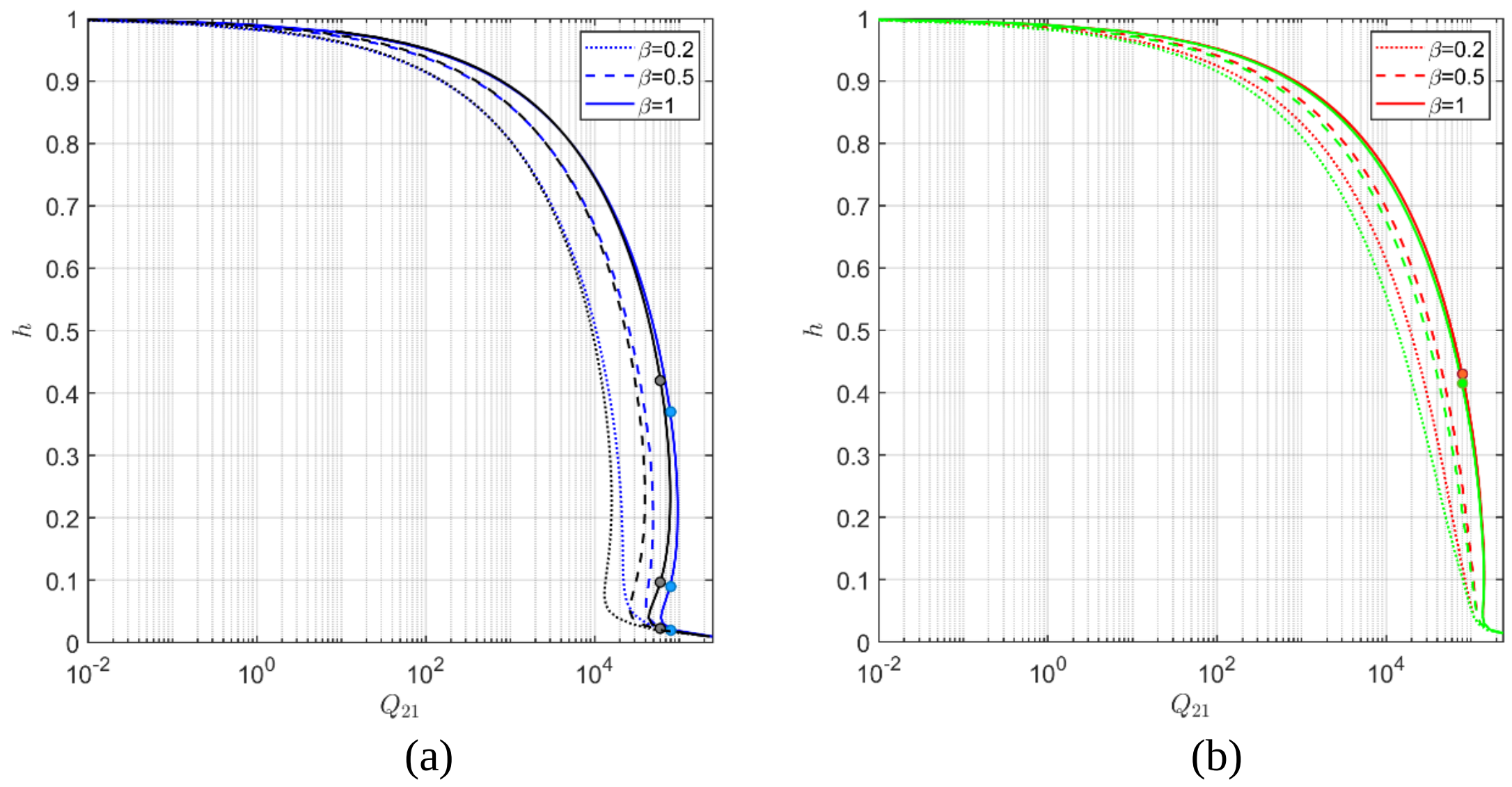

(a) (b)

**Figure 4**: Effect of the duct inclination and wall conductivities on the variation of the mercury holdup with $Q_{21}$ in concurrent upward flow, $U_{1S}$=-0.005 *m/s*, *Ha*=103.625 and *AR*=1 (a) IbIs vs CbCs, (b) CbCs vs CbIs. The blue, red, green and black lines represent IbIs, CbCs, CbIs and Ha=0, respectively. The dots in (a) mark triple solution values for $\beta$=1, *Ha*=0, $Q_{21}=6\times10^4$ and for $\beta=1^o$, *Ha*=103.625, $Q_{21}$=8 x 10^4 in IbIs duct. The corresponding single holdup solutions for $\beta=1^o$, $Q_{21}=8\times10^4$ in CbCs and CbIs ducts are marked by dots in (b).

Figure 4 shows that application of an external magnetic field results in increased holdup of the conductive liquid in upward flow, similarly to horizontal flow. Likewise, the holdup is lower in a fully insulating duct than in a perfectly conducting one. However, the holdup variation with $Q_{21}$ is highly sensitive even to a slight upward inclination. The gravity body force slows down the heavy liquid, so that in the absence of magentic field (Ha=0), for a shallow inclination of $\beta = 0.2°$ (solid black curve in Figure 4a), $h > 0.9$ up to $Q_{21} \approx 160$, compared to $Q_{21} \approx 0.2$ in horizontal duct. The application of magnetic field results in the Lorentz force that further increases the holdup, whereby the conductive fluid occupies most of the duct flow cross section over a wider range of gas flow rates. However, for the same $Q_{21}$, the influence of the wall conductivity on the holdup is drastically reduced compared to that observed in horizontal duct (see Figure 2a). As a result, the influence of the wall conductivity on $Q_{21}$ corresponding to prescribed holdup is also lower. For example, in horizontal flow and for $Ha = 103.625$, a holdup of 0.9 corresponds to $Q_{21}$ ≈10 and 80 for insulating and conducting duct, respectively, while for $\beta = 0.2°$ the corresponding $Q_{21}$ values are $Q_{21} \approx 154$ and 227, respectively. Overall, owing to the strong gravitational body force acting on the liquid phase, the Lorentz force has a significantly weaker influence on the holdup in concurrent gas-liquid upward flows than on horizontal flows, particularly when $Q_{21} < 1000$.

Inspection of Figure 4(a,b) shows that at higher $Q_{21} > 1000$, the holdup (and, as a consequence, the hydrostatic pressure drop) decreases sharply with increasing the air flow rate until it reaches a low holdup configuration that becomes almost insensitive to furhter increase of flow rate ratio. The sharp transition to lower holdup configuration is shifted to higher $Q_{21}$ as the inclination becomes steeper (larger $|\tilde{Y}|$) and the magnetic field increases. In the transition range of relatively high $Q_{21}$, three different solutions for the holdup may exist for the same air (and mercury) flow rates: a high holdup solution and two additional solutions with lower holdup values. In square ducts with a conducting bottom wall (CbCs or CbIs), the stronger retarding Lorentz force shifts the holdup transition to higher $Q_{21}$ compared to insulating duct, and a triple holdup solution is not obtained for the tested range of inclination parameters. Therefore, ignoring the effect of wall conductivity in this transition region can result in a very large error in the predicted holdup and in the associated flow characteristics.

The weakest influence of the magnetic field on the dependence of holdup on the flow rate ratio is observed in fully insulating duct (IbIs), where even under a relatively strong magnetic field (*Ha*=103.625 or $B_0$~0.2T for a 2-cm square duct) the holdup curve is quite similar to that obtained for $Ha = 0$. In both cases, a triple solution occurs over a comparable range of $Q_{21}$. The holdup curves obtained with conductive side walls (IbCs) are very similar to those obtained in insulating duct (not shown). At even higher air flow rates, beyond the triple solution region, only a single (low) holdup solution exists, which exhibits only weak sensitivity to the magnetic field intensity and wall conductivities.

### 4.1.1 High holdup range

The effect of the wall conductivities velocity contours in the flow cross-section for a holdup of 0.9 and *Ha*=103.625 are illustrated in Figures 5,6(a,b). Only the mercury velocity is shown, and the values are normalized by its maximal velocity which is reported above each frame (recall that $U_1$ is dimensionless as all velocities are normalized by $U_{1S}$ ). The less viscous and nonconductive air phase flows much faster than the mercury and drags it through interfacial shear. Consequently, the air velocity contours are then quite similar to those in single-phase Poiseuille flow in a high-aspect-ratio channel, *AR*=1/(1-*h*).

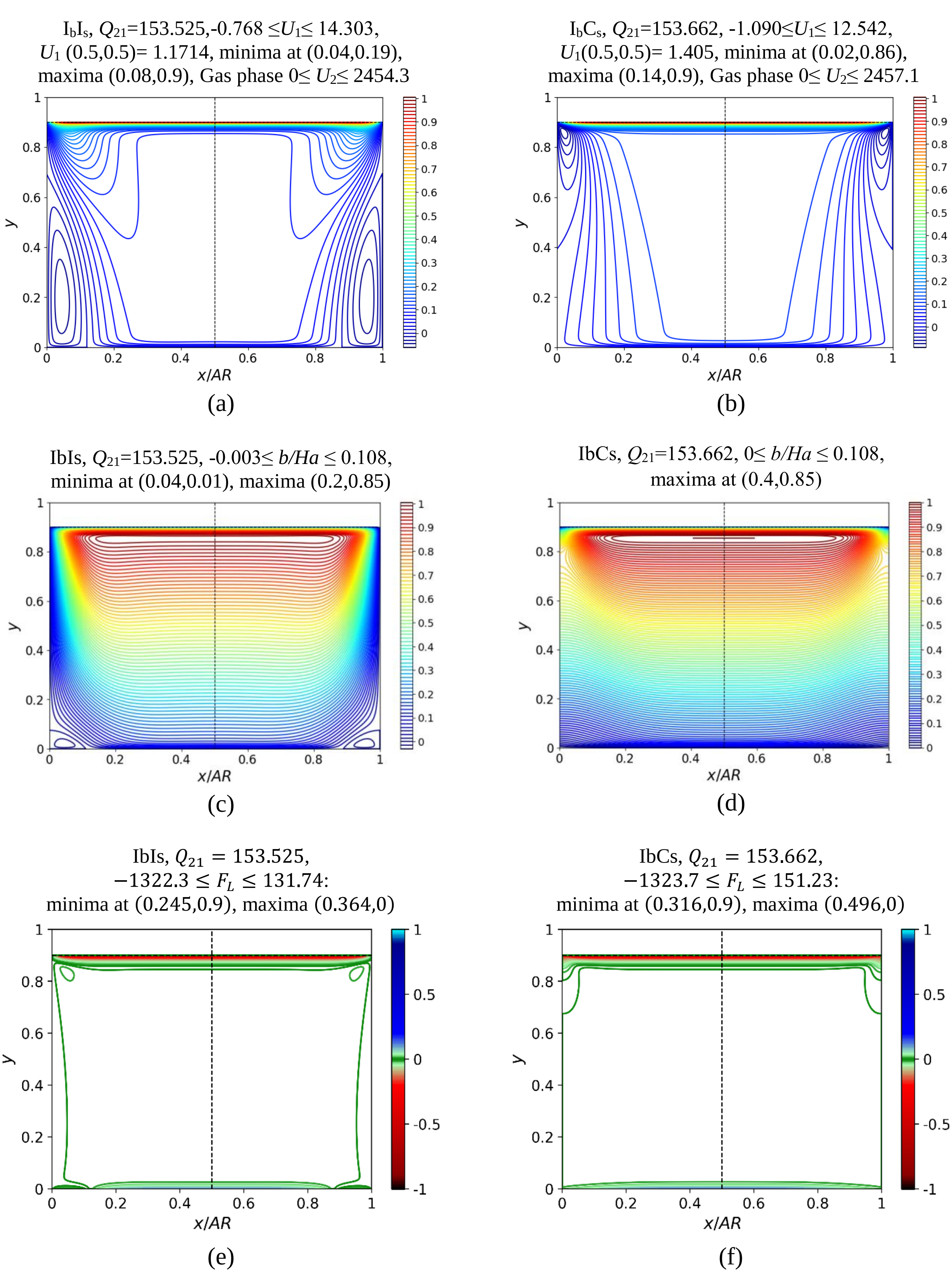

**Figure 5**: Mercury-air upward flow in IbIs (lhs frames) and IbCs (rhs frames) square ducts, $h$=0.9: (a,b) Velocity contours (c,d) induced magnetic field contours (e,f) Lorentz force distribution. ($\beta$=0.2°, $U_{1S}$ =-0.005 $m/s$, $Ha$=103.625, and $AR$=1)

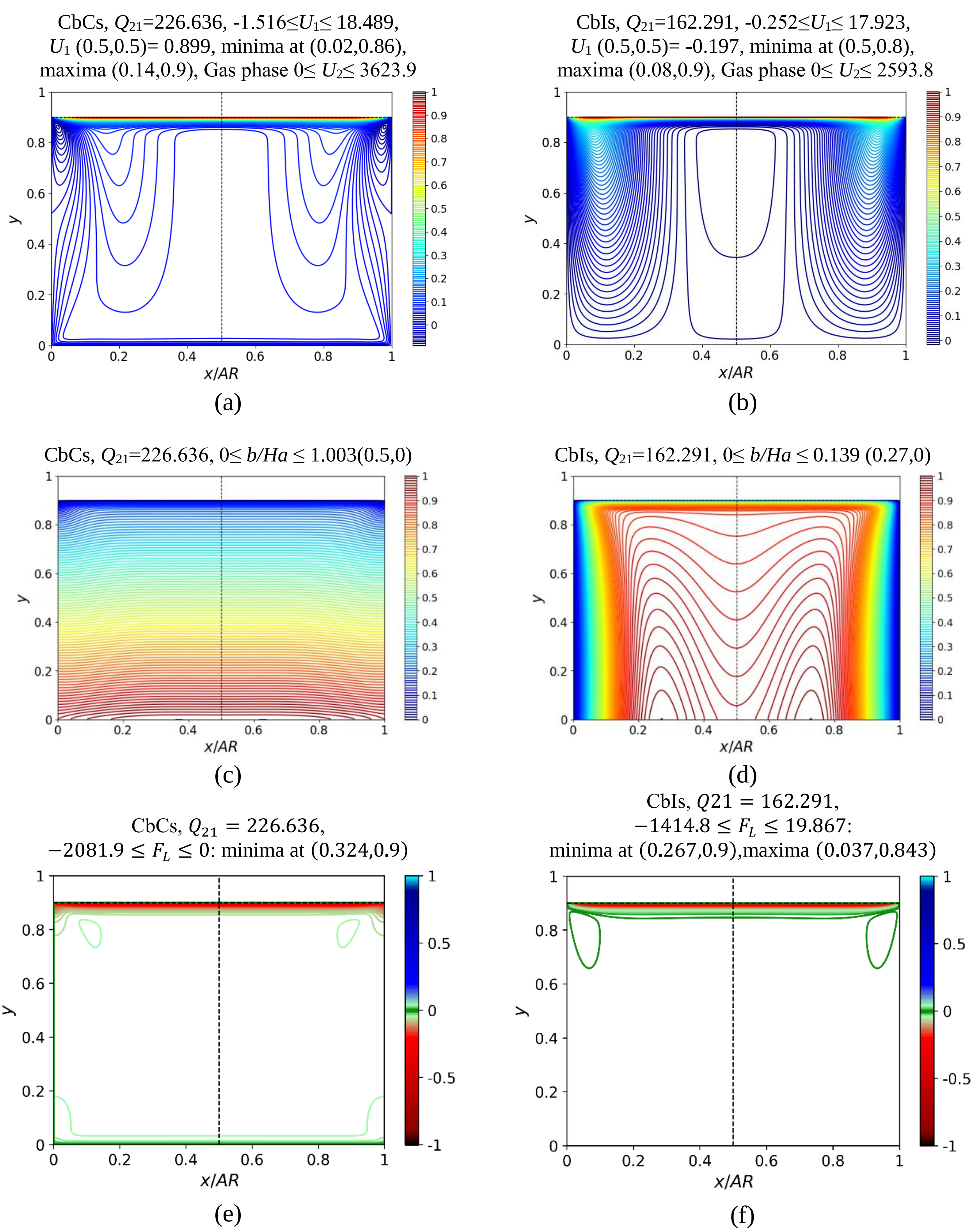

**Figure 6**: Mercury-air upward flow in CbCs (lhs frames) and CbIs (rhs frames) square ducts, $h$=0.9: (a,b) Velocity contours (b,c) induced magnetic field contours (d,e) Lorentz force distribution. ($\beta$=0.2°, $U_{1S}$ =-0.005 $m/s$, $Ha$=103.625, and $AR$=1).

A comparison of Figure 5a obtained fully insulating duct (IbIs), with Figure 5b for the duct with conducting side walls (IbCs) demonstrates the effect of replacing the insulating side walls by conducting ones. Although the velocity fields are not identical, their overall similarity is indicated by the corresponding comparable values of $Q_{21}$, as well as by the values of the maximal and minimal velocities within the flow cross section. In both cases the maximal velocity occurs at the interface, but not at its centre (unlike horizontal flow in insulating duct, see Figure 6 Subham et al., 2026 for $h$=0.9 but $Q_{21}$=1). The maximum velocity is located close to the side walls and its magnitude exceeds $U_{1s}$. by more than an order of magnitude. A large portion of the flow cross section is occupied by almost a uniform velocity region, where the velocity is of the order of $, U_{1s}$ , and the thin boundary layers near the bottom wall and the interface are clearly visible. Furthermore, the centre line velocity remains positive throughout the entire mercury layer height in both duct configurations. The gravity driven backflow regions does not appear near the bottom wall, as predicted by the TP model, but rather near the side walls. In the IbIs duct the maximal backflow is shifted toward the bottom wall, whereas in the IbCs duct the back flow is more intense, but occupies a smaller region which is closer to the interface. These backflow regions constitute the primary difference between the velocity fields in horizontal and the upward inclined ducts. They result not only from the gravitational body force but also from differences in the induced magnetic fields and the corresponding Lorentz force distributions. Identifying the intensity and location of the backflow region in the flow cross section is of particular importance, since the backflow introduces significant disturbances near the inlet, where the fluids enter the duct. It may therefore act as a source of instability for the stratified flow configuration.

The contours of the induced magnetic field, $b/Ha = B_{ind}/Re_m$ in the (IbIs) and (IbCs) ducts, corresponding to the velocity profiles in Figures 5a,b, are presented in Figure 5(c,d). The values shown are normalized by $b_{max}/Ha$. Since the induced magnetic field is zero at the wall and at the interface in these configuration, one might expect a distribution similar to that observed in single-phase flow through an insulating duct. However, unlike single-phase flow where $b_{max}/Ha$ is located at the centre of the duct regardless of the magnetic field strength, this is not the case in gas-liquid flow. In both the (IbIs) and (IbCs) ducts (Figures 5 (c,d)) the location of the high induced magnetic field values is strongly skewed towards the interface, elongated in the span-wise direction exhibiting two off-centre maxima. The direction of the Lorentz force, $F_L$= $Ha\ \partial b/\partial y = (Ha^2 Re_m^{-1}\ \partial B_{ind}/\partial y)$ is determined by the sign of $\partial b/\partial y$, and obviously varies across the conductive phase (see Figures 5e,f). As the maximum values of $b$ are close to the interface, $\partial b/\partial y > 0$ over a large portion of the conductive liquid flow cross section, where the Lorentz body force acts in the flow direction and together with the pressure gradient balances the gravity and viscous shear stresses. A region of $\partial b/\partial y < 0$ exists only in a limited region adjacent to the interface. In this region the sharp $\partial b/\partial y < 0$ results in strong retarding Lorentz forces opposing the interfacial shear exerted by the air flow. This behaviour differs from that observed in horizontal flow within the same duct, where at the same Ha, the location of $b_{max}$ is close to the bottom wall, whereby $\partial b/\partial y < 0$ throughout most of the liquid flow cross section (see Subham et al., 2026), and the Lorentz force oppose to the flow and is balanced by the pressure gradient. Near the side walls, however, $\partial b/\partial y$ is close to zero and the Lorentz force (Figure 5e,f) is ineffective in assisting the pressure gradient set in the flow in balancing the gravity and viscous shear stresses in the upward inclined ducts. These are the regions where backflow is observed in Figures 5(a,b), and are associated with reversed wall shear stress at the side walls .

Figure 6 presents the contours of the velocity, induced magnetic field, and Lorentz force obtained under the same conditions as those considered in Figure 5, but for ducts with a conducting bottom wall combined with either conducting side walls (CbCs) or insulating side walls (CbIs). The mercury velocity contours in the CbCs duct, although corresponding to a significantly higher $Q_{21}$,, exhibit some similarity to those obtained in the IbIs and IbCs ducts. In this case as well, two off-center maxima of the mercury velocity are located at the interface, although they are positioned somewhat farther from the side walls and attain higher values than in the IbIs configuration. On the other hand, the velocity in the core region is lower, and the backflow adjacent to the side walls is also more pronounced. These features result from the very different induced magnetic field obtained in the CbCs duct (Figure 6c) compared with that in IbIs duct (Figure 5c).

In the CbCs duct, the interface constitutes the only non-conducting boundary of the electrically conductive liquid. Since electric currents cannot cross this interface, they close through the low-resistance paths provided by the conducting walls, producing magnetic field lines that are nearly perpendicular to these boundaries. The resulting contour pattern is analogous to that of single-phase flow in the lower half of a duct with conducting walls, where symmetry requires that ($b$ = 0) along the horizontal centerline. The induced magnetic field intensity, (*b/Ha*) reaches its maximum value (approximately 1) at the bottom wall and decreases monotonically to zero at the insulating interface, with an almost uniform negative gradient ($\partial b/\partial y < 0$) in the span-wise direction. Consequently, the Lorentz force (Figure 6e) acts as a retarding body force, together with gravity, throughout the entire cross-section of the conductive liquid. This leads to lower core velocities in the CbCs duct, despite the higher-pressure gradient which is required to drive the flow (see Figure 11a below). In the backflow regions near the side walls, the slightly stronger Lorentz force is balanced by the reversed viscous shear stresses.

Comparison of Figures 6a (CbCs) and 6b (CbIs) demonstrates that, when the bottom wall is conducting, the side-wall conductivity has a pronounced effect on the velocity field and other flow characteristics. This influence is already evident from the substantial differences between the holdup curves corresponding to these two wall-conductivity configurations (Figure 3). It is further illustrated in Figures 6a and 6b, where, for the same holdup ($h$ = 0.9), the value of the $Q_{21}$ in the CbIs duct is approximately 25% lower than that in the CbCs duct, although it remains higher than those obtained in the in IbIs and IbCs ducts. In the CbIs duct, two high-velocity off-center regions are obtained, extending up to the interface where the maximal velocity value is attained. The upward liquid flow is transported downstream through these two regions, whereas the backflow occupies the central core of the conductive liquid ($0.4 < x < 0.6$) rather than the regions adjacent to the side walls. This backflow region extends down to the bottom wall, where the wall-shear stress reverses direction. The larger extent of the backflow region in the CbIs duct results in a reduction of the maximum backflow velocity compared with the CbCs configuration. It is worth noting that the velocity contour pattern in the CbIs duct exhibits similarities to that obtained for horizontal flow in the same duct geometry (Figure 8c in Subham et al., 2026). However, owing to the additional gravitational body force, the backflow observed in the upward-inclined duct is considerably more intense.

The magnitude and spatial distribution of the induced magnetic field (Figure 6d) and the corresponding Lorentz force contours (Figure 6f) also differ significantly from those observed in the CbCs duct (Figures 6c and 6e, respectively). In particular, the maximum value of the induced magnetic field, attained at the bottom wall, is reduced by nearly an order of magnitude. Moreover, this nearly uniform low field intensity extends over most of the conductive liquid cross-section, resulting in a weak Lorentz force in the flow direction throughout a large portion of the flow domain. The induced magnetic field decreases to zero at the insulating interface within a relatively thin layer adjacent to the interface. Consequently, steep magnetic field gradients develop in this region, giving rise to a strong retarding Lorentz force in the vicinity of the interface.

### 4.1.2 Triple holdup solutions region

As demonstrated in Figure 4, up to three distinct holdup solutions may exist within a certain range of high $Q_{21}$. The occurrence of multiple holdup solutions is characteristic of gravity-dominated two-phase flows, in which the body force is comparable to, or exceeds, the frictional pressure gradient. Under such conditions, the force balance can be satisfied by more than one flow configuration. Generally, for a given value of $\tilde{Y}$ , the range of gas flow rates over which triple solutions occur generally shifts toward higher $Q_{21}$ values and becomes narrower as the magnetic field strength increases. However, for

mercury–air flow in an insulating square duct, triple solutions persist for $\beta$ >0.5° even at *Ha*=103.625 and over a $Q_{21}$ range comparable to that observed for *Ha*=0. Such triple solutions are indicated by circles in Figure 4 for $\beta$ =1°.

The velocity contours of the triple holdup solutions obtained for *Ha*=0 and *Ha*=103.625 (IbIs duct) are shown in Figure 7(a-c) and Figure 8(a-c), respectively. It should be noted that in all cases the air velocity (not shown) is much higher than the mercury velocity and largely determines the values of the pressure gradient and the interfacial shear stress. Both attain their highest values in the high holdup solution, where the same air flow rate passes through a smaller gas flow cross section area. To clearly visualize the details of the mercury velocity field in the three configurations, the *y*-coordinate in these figures is normalized by the corresponding holdup.

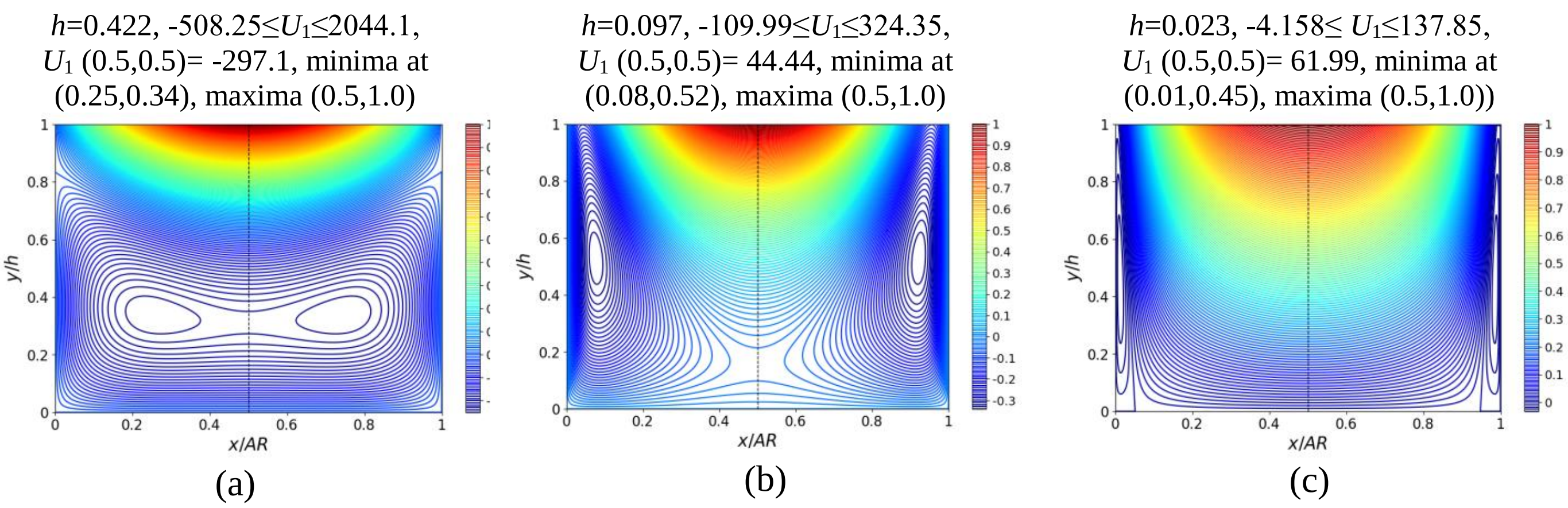

**Figure 7**: Velocity contours of the triple solutions in concurrent upward flow obtained for *Ha*=0, $U_{1S}$ =-0.005 *m/s*, $Q_{21}$=6×10$^4$, *AR*=1, $\beta$=1° (see figure 4a).

Figure 7(a–c) presents the mercury velocity fields corresponding to the three distinct holdup solutions in the absence of a magnetic field (*h* = 0.422, 0.097, and 0.023, respectively). In all the three configurations the maximal mercury velocity is at the middle of the interface. In the high-holdup solution, the mercury is subjected to the strongest downward body force due to gravity. Under these conditions, the pressure gradient and the interfacial shear exerted by the air are insufficient to transport the entire mercury layer upward. As a result, a pronounced mercury backflow develops, reaching its maximum magnitude at two symmetric off-center locations. This backflow occupies a substantial portion of the cross section and result in reversed wall shear over the bottom wall and a large part of the side walls. In the intermediate-holdup solution, virtually no backflow is observed adjacent to the bottom wall. Instead, the backflow regions are displaced toward the side walls, and their intensity is significantly reduced. In the low-holdup solution, the gravitational force acting on the thin mercury layer is much weaker. Consequently, the pressure gradient and the interfacial shear are sufficient to drive upward flow through almost the entire mercury layer. Mercury backflow is then confined to very narrow regions adjacent to the side walls.

The triple solutions shown in Figure 8a-c ($Ha$=103.325, IbIs duct) correspond to holdups similar to those obtained for Ha=0, but for approximately 30% higher air flow rate. When a magnetic field is applied, the resulting Lorentz force depends on the distribution of the induced magnetic field (Figure 9 a-c) and may either enhance or counteract the opposing gravitational body force (see Figure 10a–c).

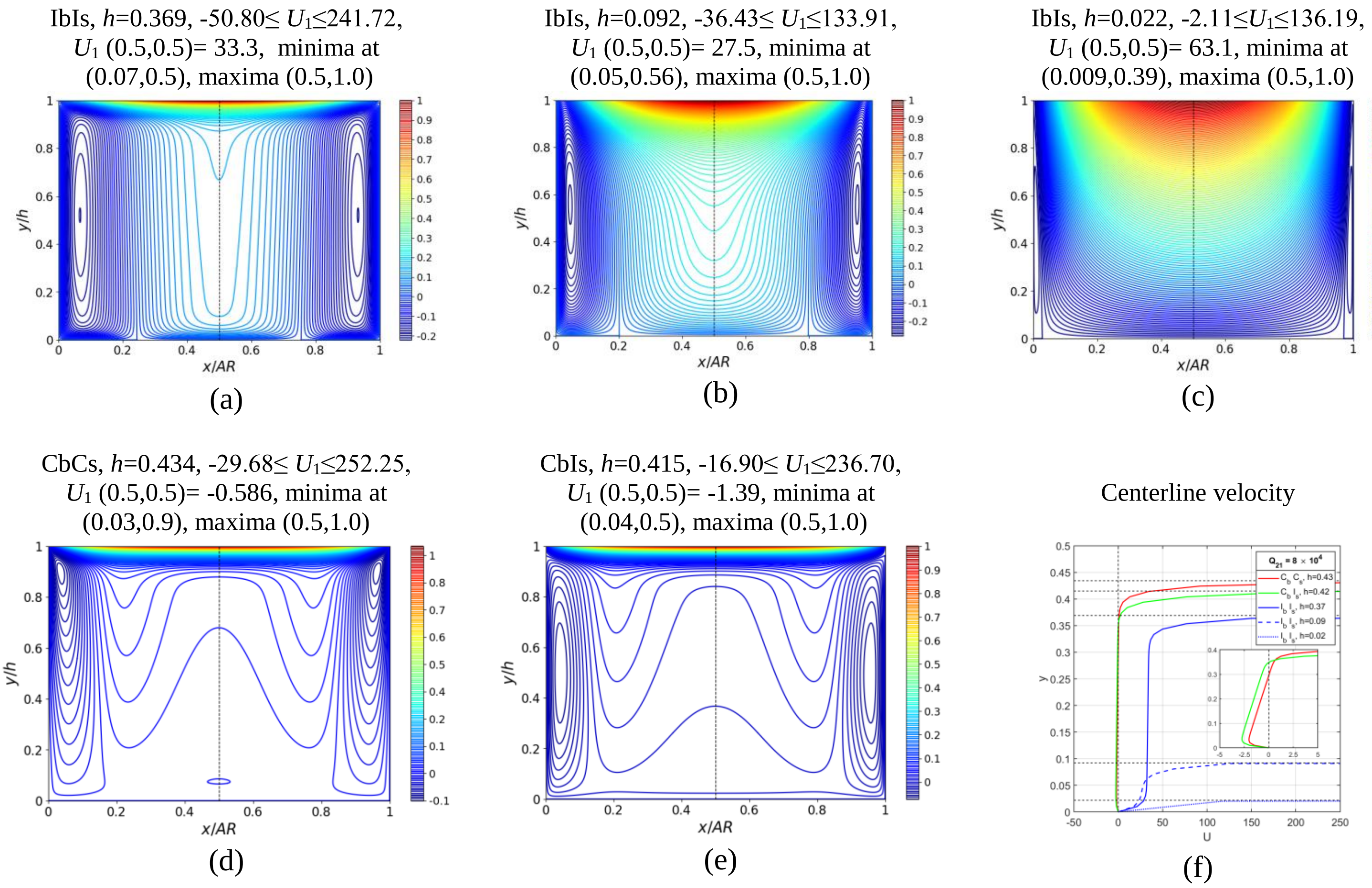

**Figure 8**: Velocity contours for the cases indicated by dots in Figure 4 ($U_{1S}$ =-0.005 $m/s$, $Q_{21}$=8×10$^4$, $AR$=1, $\beta$=1°, $Ha$=103.625) (a-c) The triple solution obtained in IbIs duct, (d) and (e) the single solution for CbCs and CbIs ducts. (f) The corresponding centerline velocity profiles.

Similar to Figures 5c, in the upper- and middle-holdup solutions, the maximum of b/Ha is skewed toward the interface. However, its value in Figure 8c is considerably higher, resulting in stronger Lorentz forces. Here too, the Lorentz forces acting in the flow direction and opposing gravity extend over a large portion of the mercury core where ∂b/∂y > 0, with the highest values occurring near the bottom wall (blue region in Figures 10a,b). The highest retarding Lorentz forces are near the interface (red region in Figures 10a,b) and also observed in the centerline Lorenz force profile (Figure 10f). Also noteworthy are the negative values of b (i.e., directed opposite to the local flow) and the corresponding negative Lorentz forces acting also near the lower part of the side walls. In these regions the Lorentz forces enhance the effect of the retarding gravity force and result in the backflow regions near the side walls observed in Figures 8a,b. In the lower solution, the $b$/Ha-contours (Figure 9c) and the corresponding Lorentz force distribution (Figure 10c) are similar to those obtained in shear dominated single-phase flow of a conductive liquid in a rectangular duct of an aspect ratio $AR=h$, where the $b_{max}$ is located at the duct centre.

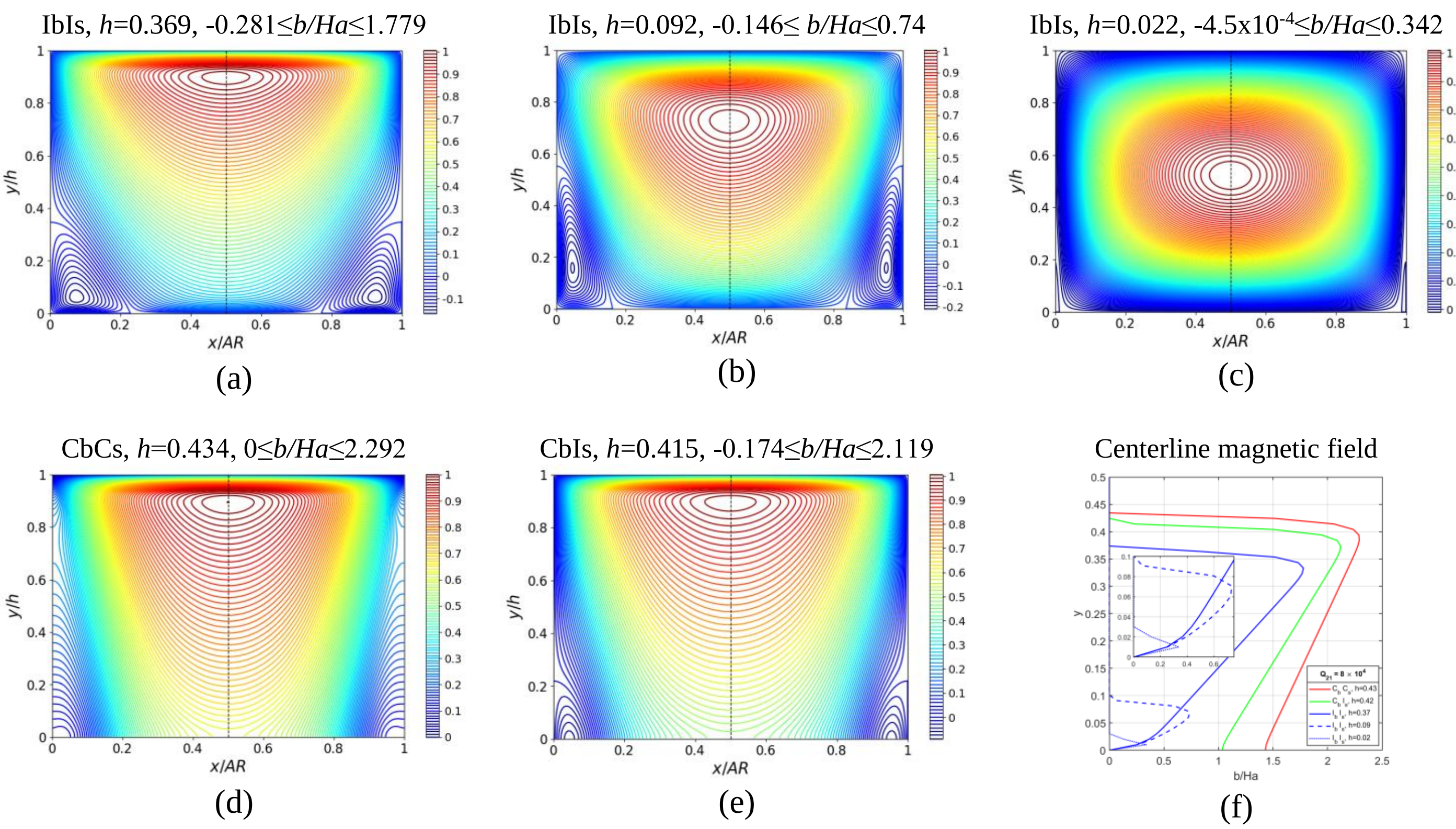

**Figure 9**: (a-e) Induced magnetic field corresponding to the velocity contours depicted in Figure 8 (a-e), (f) the centerline profiles ($U_{1S}$ =-0.005 $m/s$, $Q_{21}$=8×10$^4$, $AR$=1, $\beta$=1° and $Ha$=103.625

Comparison of velocity contours in Figure 7(a-c) with the corresponding holdups in Figure 8(a-c) reveals the overall impact of the Lorentz force in weakening the influence of gravity on the velocity field. In all three solutions the backflow intensity is significantly reduced compared with the corresponding $Ha$ = 0 cases. In particular, for the high holdup solution, the maximal backflow velocity is reduced by an order of magnitude, while the backflow region becomes considerably smaller. Similar to the middle-holdup solution, the reverse flow is confined to two regions adjacent to the side walls, with no backflow present in the duct core. The velocity field of the lower-holdup solution, which is dominated by the interfacial shear exerted by the gas, remains practically unchanged relative to the Ha = 0 case. This behavior is clearly reflected in the centerline velocity profiles corresponding to the three solutions, shown in Figure 9c.

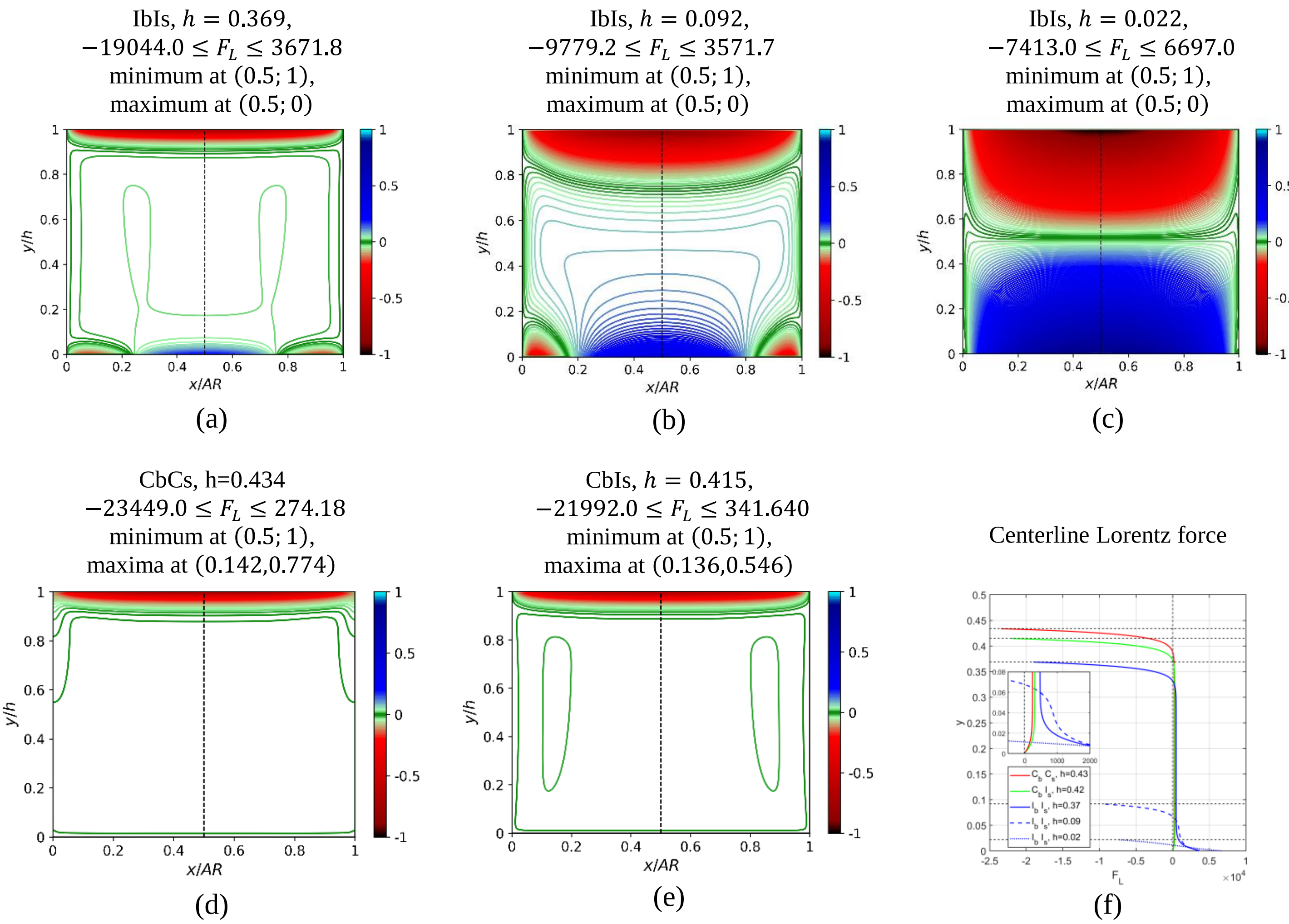

**Figure 10**: (a-e) The Lorentz force distribution corresponding to the velocity contours depicted in Figure 8 Contour plots of $F_L$ scaled by its maximal absolute value, i.e., $F_L/\max(abs(F_L))$. (f) the centerline profiles ($U_{1S}$ =-0.005 *m/s*, $Q_{21}$=8×10$^4$, *AR*=1, *β*=1° and *Ha*=103.625).

In ducts with conducting bottom wall (CbCs and CbIs), triple-holdups are not obtained (see Figure 4). For the same $Q_{21}$ (=8×10$^4$) examined in the IbIs duct (Figure 8), the holdup in the CbCs and CbIs ducts (*h*=0.434 and 0.415, respectively) is higher than that of the upper holdup solution in the IbIs duct (*h*=0.369). The corresponding velocity fields (Figure 8d,e) indicate that, in the high air flow rates region, both the flow structure and the Lorentz force distribution (Figures 10d,e) resulting from the induced magnetic field (Figures 9d,e) are similar in the CbCs and CbIs ducts. This similarity is also indicated by the corresponding centreline profiles of the induced magnetic field and Lorentz force (Figures 9f and 10f). Compared to the upper holdup solution in IbIs duct (Figure 10a), the retarding Lorentz force is significantly stronger in the CbCs and CbIs, whereas the maximal positive Lorentz force (opposing gravity) is approximately one order of magnitude weaker. Although the latter is significantly higher than that obtained in these ducts in the high holdup region (Figure 6e,f, $h$=0.9), it is insufficient to supress mercury backflow. Compared to the velocity contours obtained in the CbCs and CbIs ducts in the high holdup region (Figures 6a,b), the maximum velocity in Figures 8(d,c) is located at the centre of the interface and is substantially higher, whereas the backflow near the side walls is intensified. Backflow is present also in the lower part of the mercury core (see Figure 8f), where the Lorentz force and pressure gradient are insufficient to overcome the gravitational body force.

#### 4.2.3 Pressure gradient and pumping power

Figure 11 shows that the combined effects of wall conductivities on the conductive phase holdup, the Lorentz force and the velocity profiles lead to different values of the pressure gradient required to sustain specified fluid flow rates in ducts with different wall conductivities. The $\tilde{P}_f^0$, $\tilde{P}_g^0$ (= $(\tilde{P}_f, \tilde{P}_g)/28.455$, shown in Figure 11(a,b), represent the frictional and gravitational pressure gradient, respectively, normalized by the frictional pressure gradient in single-phase flow of the conductive liquid in a square duct without an applied magnetic field (Ha=0). The frictional pressure gradient for $Ha$=0 and $\beta = 1^{\circ}$ is also included in Figure 11a as a reference highlighting the effect of the applied magnetic field (Ha=103.625). Note that for $Ha$=0, $\tilde{P}_f^0 \rightarrow 1$ as $Q_{21} \rightarrow 0$ and in upward inclined flow it monotonically increases with increasing the air flow rate.

When a magnetic field is applied, the frictional pressure gradient represents the combined effects of the viscous shear and Lorentz force acting in the flow cross section. As expected, the application of a magnetic field increases the frictional pressure gradient. However, comparison of the $\tilde{P}_f^0$ values obtained for $Ha$=103.625 and $Ha$=0 indicates that in a fully insulating (IbIs) duct the increase of $\tilde{P}_f^0$ with Ha is relatively modest. Furthermore, the variation of $\tilde{P}_f^0$ with $Q_{21}$ follows a similar trend to that obtained for $Ha$=0 already for low $Q_{21}$ . This is attributed to the lower Lorentz force in the IbIs duct, which is also reflected in highest sensitivity of $\tilde{P}_f^0$ to increase of the duct inclination observed already at low $Q_{21}$ values. Similar $\tilde{P}_f^0$ are obtained for the IbCs (not shown). The highest $\tilde{P}_f^0$ values are obtained in the fully conducting (CbCs) duct, where the dominance of the Lorentz force over gravity results in insensitivity of the frictional pressure gradient to the duct inclination over the widest range of $Q_{21}$. Replacing the conducting side walls with insulating walls (CbIs) leads to a drastic reduction in the frictional pressure gradient (by a factor of about 6.5), resulting in values closer to those in the IbIs (and IbCs) duct.

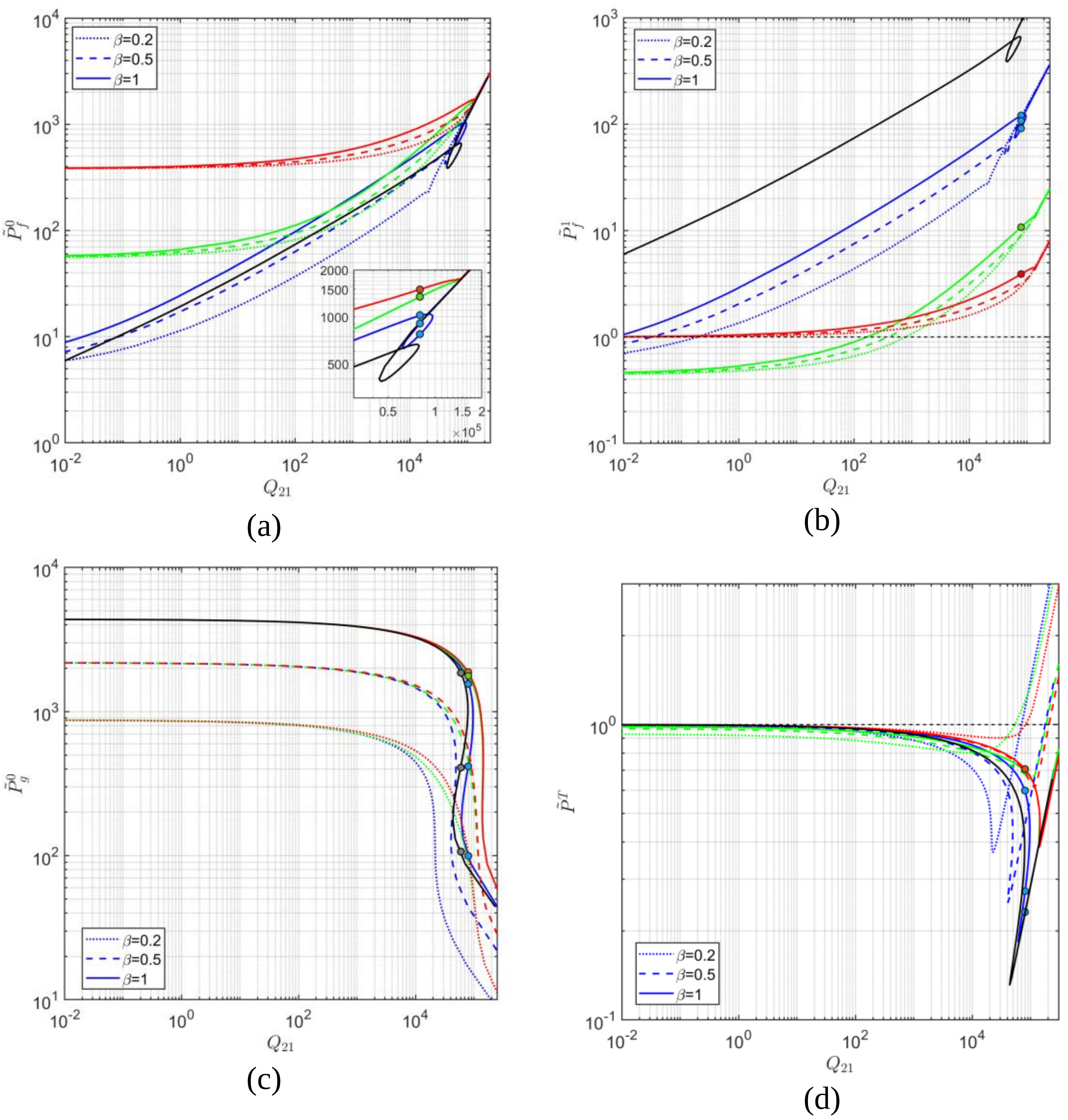

**Figure 11**: Variation of the pressure gradient with $Q_{21}$ in concurrent upward flow: Effect of the duct inclination and wall conductivities ($U_{1S}$ =-0.005 *m/s*, *Ha*=103.625 and *AR*=1) (a) $\tilde{P}_f^0$, (b) $\tilde{P}_f^1$, (c) $\tilde{P}_g^0$ (d) $\tilde{P}^{1T}$. The blue, red, green and black lines represent IbIs, CbCs, CbIs and Ha=0, respectively. The dots mark the triple solution obtained in IbIs duct for $Q_{21}=8\times10^4$ and the corresponding single solution in CbCs and CbIs ducts

For high $Q_{21}$, the sensitivity of the frictional pressure gradient to the wall conductivity configuration decreases. The triple values of the frictional pressure gradient obtained for the IbIs duct correspond to the triple holdup solutions identified in Figure 4. The lowest frictional pressure gradient is associated with the lower-holdup solution, and the highest one is similar to that obtained in the (CbCs) and (CbIs) ducts. At still higher $Q_{21}$, where the liquid flow becomes dominated by interfacial shear and only a single solution exists for all duct configurations, the frictional pressure gradient becomes essentially independent of the magnetic field intensity and the wall conductivities and increases monotonically with air flow rate.

The possibility of achieving a lubrication effect by air injection is examined in Figure 11b, presenting the frictional pressure gradient factor, $\tilde{P}_f^1$ (i.e., the frictional pressure gradient normalized by that obtained in single-phase mercury flow in the same duct and under the same magnetic field strength, $\tilde{P}_f^1 = \tilde{P}_f^0/\tilde{P}_{f\,1s}^0$). In all cases $\tilde{P}_f^1$ approaches unity as $Q_{21} \to 0$ (out of the range shown in the figure). The $\tilde{P}_{f_{1s}}^0$ values for single-phase mercury flow at *Ha*=103.625 in the various duct configurations are 8.429, 8.555, 384.01,124.51 in IbIs, IbCs, Cb,Cs, CbIs ducts, respectively.

Figure 11b shows that for Ha=0, $\tilde{P}_f^1 = \tilde{P}_f^0$ increases with $Q_{21}$, hence, no lubrication effect is obtained by introducing air flow to the mercury flow in an upward inclined duct. The results of $\tilde{P}_f^1$ for *Ha*= 103.625 indicate that the application of external magnetic field reduces the $\tilde{P}_f^1$ values. However, in fully conducting duct and, $\tilde{P}_f^1$ remains >1 throughout the entire $Q_{21} > 0$ range, and increases with increasing the air flow rate. In insulating duct (IbIs) some lubrication can be obtained at low air flow rate and $\beta$ <1$^o$ . The most significant lubrication effect is obtained in the CbIs duct, where $\tilde{P}_f^1 < 1$ up to $Q_{21} = 100$ even for $\beta$ =1$^o$. For $0.01 < Q_{21} < 1, \tilde{P}_f^1 \approx 0.5$ (i.e., 50% reduction of the frictional pressure gradient) with low sensitivity to the duct inclination. The lubrication achieved is attributed to the extensive backflow in the CbIs duct, which extends down to the bottom wall and the results in a reversed wall shear (see Figure 6b), as well as the lower Lorentz force (see Figure 6f) affected by introducing the air flow.

The variation of the hydrostatic pressure gradient factor, $\tilde{P}_g^0$ with $Q_{21}$ is shown in Figure 11c. Since the contribution of the air phase to the hydrostatic pressure gradient is negligible, the $\tilde{P}_g^0$follows the same trends of the mercury holdup shown in Figure 4. Consequently, the significant effect of wall conductivities observed in Figure 11c at high $Q_{21}$ >$\sim 1000$ , where the lowest $\tilde{P}_g^0$values are obtained for IbIs duct is expected. However, comparison of $\tilde{P}_g^0$ with the corresponding $\tilde{P}_f^0$ values in Figure 11a indicates that for all duct configurations, $\tilde{P}_g^0$ constitutes a significant contribution to the total pressure gradient over a wide range of $Q_{21}$ even at relatively small upward inclination of $\beta$ =0.2$^o$., except for the low holdups obtained beyond the triple solution region where the contribution of the hydrostatic pressure gradient is obviously low.

From the perspective of reducing the pressure gradient through air injection, the total pressure gradient factor, $\tilde{P}^T$ should be considered. This factor is defined as $dp/dz$ normalized with respect to the total pressure gradient of single-phase flow of the conductive liquid under the same magnetic field strength, see Eq. 16). Figure 11d shows that for $Q_{21}$ <~1000 , introducing the air flow practically does not affect the total pressure gradient in IbIs and CbCs ducts. Consequently, introducing air flow increases the pumping power. In CbIs duct, small reduction of the total pressure gradient is achieved by introducing air flow, with possible similar power reduction at $Q_{21} \ll 1$. However, within and near the triple solution region, the sharp reduction of the mercury holdup (and thus the hydrostatic pressure gradient), leads to a pronounced reduction of the total pressure gradient compared to single-phase mercury upward flow. The most pronounced reduction is observed in a IbIs duct (e.g., by a factor of approximately 0.4, 0.2 for $\beta$ =0.2$^o$, 1$^o$, respectively, at Ha=103.625). Nevertheless, because such reductions are achieved only at very high values of $Q_{21}$, pumping power reduction is not achievable also in this region.

### 4.2. Countercurrent and concurrent down flows in square ducts

Countercurrent gas-liquid flow is a configuration encountered in many heat and mass transfer processes. The liquid flows downward ( $U_{1s}$>0), resulting in $\tilde{Y} > 0$, whereas the gas flows upward ( $U_{2s}$ <0) yielding $Q_{21}$ <0. Countercurrent flow is a gravity dominated regime that can be established for sufficiently large inclination parameter, and in a limited range of upward gas flow rates that remain below the flooding limit.

Figure (12a) illustrates the effects of the duct inclination on the mercury holdup variation with $Q_{21}$ for an applied magnetic field corresponding to Ha=103.625. For example, $U_{1s} = 1$ m/s and β=5$^o$, corresponds to $\tilde{Y} = 3098.3$ in a 0.02m square duct. The resulting holdup curve shows the variation of the mercury holdup with the upward air flow rate. The figure also includes the corresponding holdup results for *Ha* = 0. The latter serves as a reference for assessing the effect of the magnetic field and demonstrating the significance of the wall-conductivity configuration on the mercury holdup.

As expected, the application of magnetic field weakens the effect of gravity through the Lorentz force, and reduces the $Q_{21}$ range where countercurrent flow is feasible. However, as in the concurrent upward flow, the influence of the magnetic field on the holdup curve is the weakest in an ideally insulating duct (IbIs). For this duct at $\beta$=5$^o$, the countercurrent flow region remains comparable to that obtained for *Ha*=0 even under a strong magnetic field. Although the countercurrent flow region decreases as the inclinations angle becomes shallower, it remains feasible even for $\beta$=1$^o$. Similar results are obtained in a IbCs duct (not shown). However, when the bottom wall is conducting (CbCs or CbIs), no countercurrent region exists even at $\beta$=5$^o$ owing to the stronger retarding Lorentz force. Countercurrent flow can be restored in these ducts when the mercury flow rate is reduced (i.e., increasing $\tilde{Y}$ ). This is demonstrated in Figure 12b by decreasing $U_{1s}$ from 1 m/s to 0.1 m/s, thereby increasing the value of $\tilde{Y}$ by an order of magnitude (to $\tilde{Y} = 30983$). Under these conditions, replacing the conductive side walls in a CbCs duct with insulating ones (i.e., CbIs) extends countercurrent flow region, demonstrating the beneficial effect of insulating side walls. Nevertheless, this region remains substantially narrower than that obtained in the fully insulating duct, where the countercurrent region is even slightly wider than that obtained for *Ha*=0. This behavior results from the combination of a weak Lorentz force in insulating ducts and the enhanced influence of gravity at the higher $\tilde{Y}$ , which is amplified by the higher mercury holdups that are obtained in the presence of the magnetic field.

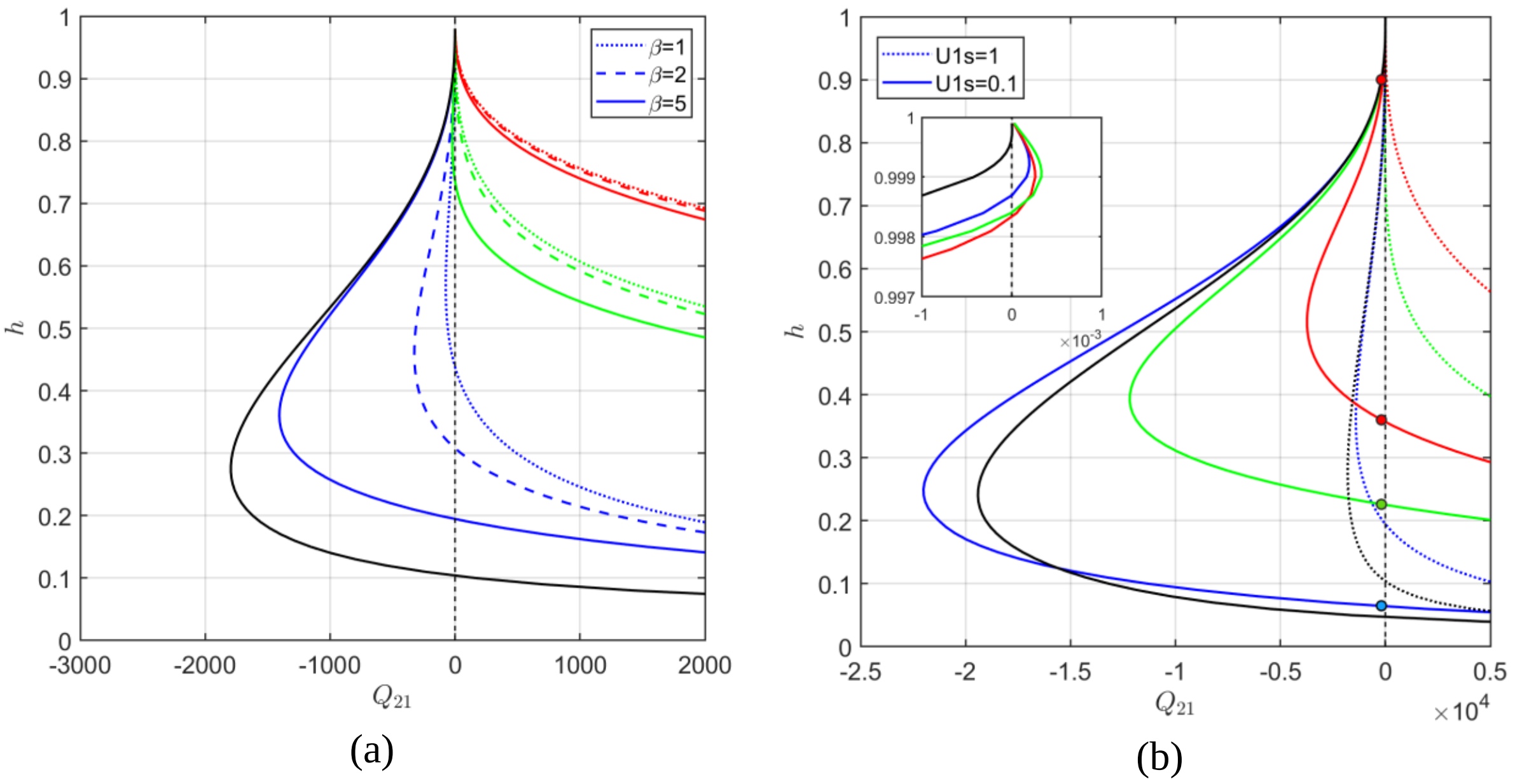

**Figure 12**: Variation of the mercury holdup with $Q_{21}$ in downward flow of mercury (*Ha*=103.625 and *AR*=1). $Q_{21}<0$ corresponds to countercurrent flow (a) Effect of the duct inclination and wall conductivities, $U_{1S}$ =1 *m/s* (b) Effect of $U_{1S}$ and wall conductivities at $\beta$=5° (note the change of scale). The blue, red, green and black lines represent IbIs, CbCs, CbIs and *Ha*=0, respectively. The dots mark double solutions obtained in countercurrent flow when the upper solution in all cases is *h*=0.9.

Figures 12 (a,b) demonstrate that within the feasible countercurrent flow region, two distinct holdups (and corresponding flow characteristics) are obtained for a given value of $Q_{21}$. The flow configuration that is realized depends on the resistance imposed at the heavy phase outlet. The high holdup configuration can be obtained by increasing the outlet resistance, while maintaining the same phases' flow rates (e.g., Ullmann et al., 2003). At the flooding point, the two solution branches merge into a single solution. Beyond this point, no steady countercurrent flow exists. In practice, for a given wall-conductivity configuration and inclination angle, if the gas flow rate exceeds the flooding point, part of the liquid must be allowed to be carried concurrently upward with the gas for preventing liquid accumulation at the inlet. Consequently, the downward liquid superficial velocity ($U_{1s}$ ) is reduced, resulting in a higher value of $\tilde{Y}$ that permits the establishment of countercurrent flow regime.

Apparently, Figures 12(a,b) suggest that in concurrent downward flow, for specified wall conductivity configuration and values of $Ha$, $\tilde{Y}$ and $Q_{21} > 0$, a single solution of a low holdup is obtained. This solution extends from the lower solution branch of the countercurrent flow region into the concurrent downward flow region through $Q_{21} = 0$. As in the countercurrent flow region, the holdup along this branch is the lowest in the IbIs duct and the highest in the CbCs duct. However, enlargement of the region characterized by high holdups and low $Q_{21}$ (see insert in Figure 12b) reveals that two additional high-holdup solutions can be obtained for sufficiently large $\tilde{Y}$. The middle holdup solution is an extension of the upper holdup solution branch of the countercurrent flow region and is obtained when this branch crosses the $Q_{21} = 0$ at a value of $h$<1. In both the middle and low holdup solutions obtained for $Q_{21} = 0$ the gas circulates in the duct, where its upward flow due to gravity is balanced by its downward flow near the interface, where the gas is dragged downward by the liquid. The upper solution for $Q_{21} = 0$ always corresponds to single phase flow of the liquid ($h$=1). The triple solution region in concurrent down flow is however limited to a very narrow range of low gas flow rates, above which only a single solution is obtained.

The effect of the wall-conductivity configuration on the velocity, induced magnetic field and Lorentz force contours within the flow cross-section is illustrated in Figures 13 to 15 for the dual hold solutions obtained for $U_{1s}$=0.1, β=5$^o$ and $Ha$=103.625. In all cases the high holdup is fixed at $h$=0.9 (l.h.s frames). The corresponding value of $Q_{21} < 0$ for all wall conductivity configurations. However, its value as well as the associated low holdup solution (r.h.s frames) varies with the wall conductivity configuration. These figures provide insight into the mechanisms by which the wall-conductivity arrangement influences the flow field, induced magnetic field, and Lorentz-force distribution, thereby affecting the characteristics of the dual-holdup solutions.

Figures 13(a,b) present the axial velocity contours of the dual flow configurations obtained in IbIs. In the high holdup configuration (Figures 13a), the liquid core flows downward ($U_{1S}$>0), and is of the order of $U_{1S}$. However, near the interface, the strong interfacial shear generated by the gas upward flow through the narrow gas gap, entrains the liquid upward, producing a liquid backflow region (i.e., local negative velocities). The maximal liquid velocity within the flow cross section is located in two-off-center regions close to the side walls. In fact, the dimensionless velocity contours in the high holdup configuration of countercurrent flow resemble those obtained for the same holdup in concurrent upward flow for the same $Ha$ (e.g., compare Figures 12a with Figures 5a obtained for upward flow in IbIs for $h$=0.9). As expected, the liquid backflow regions in countercurrent flow correspond to the regions of high liquid velocity in concurrent upward flow, and vice versa.

In the low-holdup solution (Figures 13b), the liquid occupies a smaller fraction of the cross-section and the gas flows through a wider channel. In this configuration, the average downward velocity is obviously higher ($\overline{U}_1 = U_{1s}/h$). The downward-flowing mercury exerts sufficient interfacial shear to drag the air downward in the vicinity of the interface, creating a localized air-backflow region, which is maximal near the side walls (at $x$=0.11). In this configuration, no mercury backflow is observed in the flow cross section. The highest mercury velocities occur near the interface, where they reach values almost twenty times greater than $U_{1S}$.

As in concurrent upward flow, also in countercurrent flow, the conductivity of the side walls has only a minor effect on the flow characteristics when the bottom wall is insulating, both for the low and high holdup solutions. This is a result of the nearly identical induced magnetic field distributions (shown in Figure 13c,d) and the corresponding Lorentz-force distributions (Figure 13e,f) within the flow cross-section (results for IbCs duct are therefore not shown). In both wall conductivity configurations, in the high-holdup solution, the mercury experiences a relatively low retarding Lorentz force ($F_L$<0) over most of the flow cross-section, and a strong Lorentz force assisting the flow near the interface ($F_L$>0, blue color in Figure 13e). In the low-holdup solution, the Lorentz force opposes the mercury flow in the upper part of the mercury layer (red color in Figure 13f) and assists the flow in its lower part.

Figures 14(a–d) presents the axial velocity contours of the dual flow configurations obtained when the bottom wall is conducting, namely in the CbCs and CbIs ducts. In the CbCs duct, the value of $Q_{21}$ corresponding to $h$=0.9 is approximately 60% lower than that obtained in the ducts with an insulating bottom wall. Furthermore, the associated low-holdup solution is considerably thicker, with a holdup approximately 5.6 times greater than that of the corresponding solution in the insulating-bottom-wall ducts and the average velocity is therefore significantly lower. Nevertheless, the velocity contours of the high holdup configuration (Figures 14a) still resemble those obtained in the IbIs (or IbCs) duct (Figures 13a), where the air drags the mercury upward near the interface, and the highest mercury velocity is observed in two off-center regions close to the side walls. Inspection of the induced magnetic field contours, which are normalized by the maximal value in the cross section (Figure 15a), shows similarity with those obtained in the IbCs duct (Figure 13c). However, the maximal *b/Ha* is 8 times larger in the CbCs duct, resulting in a stronger retarding Lorentz force

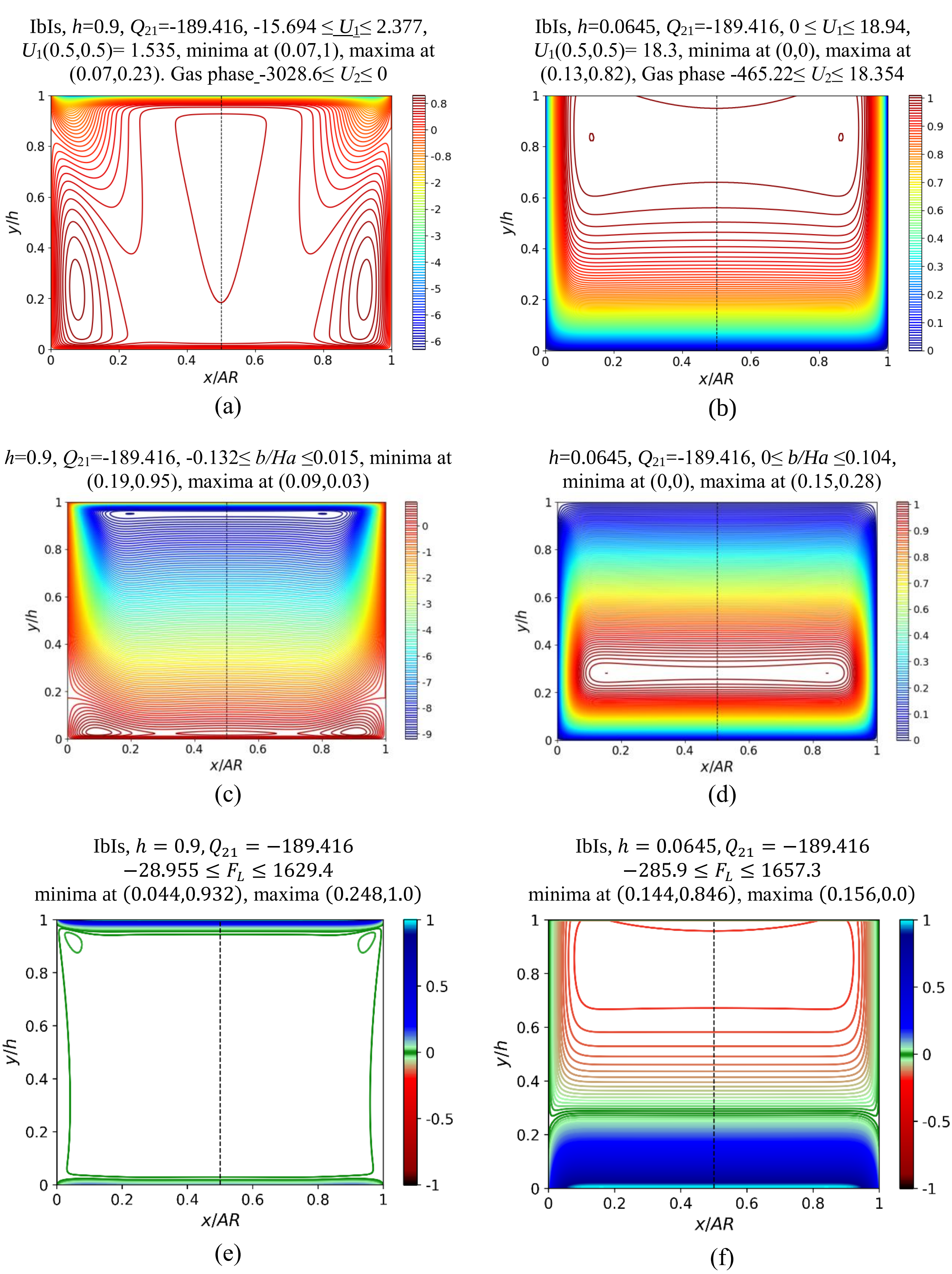

**Figure 13**: Dual solutions in countercurrent flow in a IbIs duct. L.h.s frames correspond to the upper solution ($h$=0.9), r.h.s. frames to the lower solution ($h$=0.06). (a,b) Velocity contours (c,d) Induced magnetic field contours (e,f) Lorentz force contours ($U_{1S}$ =0.1 $m/s$, $Ha$=103.625, $\beta$=5°, $AR$=1).Contours of the force normalized by its maximal absolute value, i.e., $F_L/\max(abs(F_L)$.

experienced by the mercury core (compare Figure 16a to 13e), and therefore, its velocity is lower compared to that obtained in the IbIs duct (compare 14a and 13a). The mercury backflow induced by the upward air shear at the interface must be reduced in the CbCs duct in order to satisfy the prescribed value of $U_{1S}$. Consequently, to maintain the same holdup ($h$ = 0.9), a lower air countercurrent flow rate (i.e., lower $Q_{21}$) is required in the CbCs duct than in the IbIs and IbCs ducts.

Examining the induced magnetic field contours of the dual solutions obtained in the CbCs duct (Figures 15a,b) reveals that unlike the IbIs and IbCs ducts, in the CbCs duct, the induced magnetic field contours are similar with practically the same maximal *b/Ha* value. In both the high and low holdup configurations, a retarding Lorentz force acts over most of the mercury flow cross-section (Figures 16a,b). However, in the low-holdup configuration, the Lorentz force decreases from its maximum value at the conducting bottom wall to zero at the interface over a thinner mercury layer. As a result, the mercury core is subjected to a stronger average retarding Lorentz force than in the high-holdup configuration, which significantly modifies the velocity field. In the low-holdup configuration (Figure 15b), the regions of highest velocity are located along the side walls. Although the location of the maximum velocity is similar to that observed in the low-holdup solutions of the IbIs (or IbCs) duct (Figure 13b), its magnitude is approximately 5.5 times lower, and the shape of the velocity contours is significantly different, reflecting the stronger influence of the Lorentz force in the CbCs configuration.

Comparison of the velocity contours obtained in the CbCs and CbIs ducts for the high-holdup configuration (Figures 14a and 14c) and the low-holdup configuration (Figures 14b and 14d) reveals the pronounced influence of side-wall conductivity when the bottom wall is conducting. This effect is already reflected in the higher value of $Q_{21}$and the lower holdup of the low-holdup solution in the CbIs duct, which approaches the value obtained in the IbIs (or IbCs) duct (Figure 13b). In the high-holdup configuration, the mercury backflow near the interface resulting from the upward gas-flow shear, is reduced. The maximum velocity is located closer to the side walls and is more than three times higher than that in the CbCs duct, whereas the velocity in the core is approximately five times lower and amounts to only about 25% of $U_{1S}$.. As a result, the velocity field exhibits a pronounced jet-like structure, in which the local velocities in the side-wall jets exceed those in the core region. This jet-like flow pattern characterizes also the low-holdup configuration (Figure 14d). In this case, in the two high-velocity jets adjacent to the side walls the maximum velocity reaches $17.4U_{1S}$, however, it is only about 6 times higher than the core velocity (e.g., the centerline velocity at $U(y/h{=}0.5){=}2.9$). This jet-like flow structure results from the significantly stronger retarding Lorentz force acting in the mercury core compared with that in the IbIs and IbCs ducts (Figures 13e and 13f), combined with the relatively low Lorentz force near the side walls compared to those acting in the mercury core.

The variation of frictional and hydrostatic pressure gradient factors, $\tilde{P}_f^0$, $P_f^1$ and $\tilde{P}_g^0$, with $Q_{21}$ in a square duct are shown in Figures 17(a-c) for *Ha*=103.625 and $\tilde{Y} = 30983$ (corresponding to $U_{1s} = 0.1$ m/s, β=5°, *H*= 0.02m). The effect of the magnetic field and the significance of the wall-conductivity configuration on the pressure gradient factors is demonstrated by comparison with the corresponding pressure gradient factors obtained in the absence of magnetic field (*Ha*=0) under otherwise identical conditions. Note that $P_f^1 = \tilde{P}_f^0$ for *Ha*=0, therefore it is not included in Figure 17b.

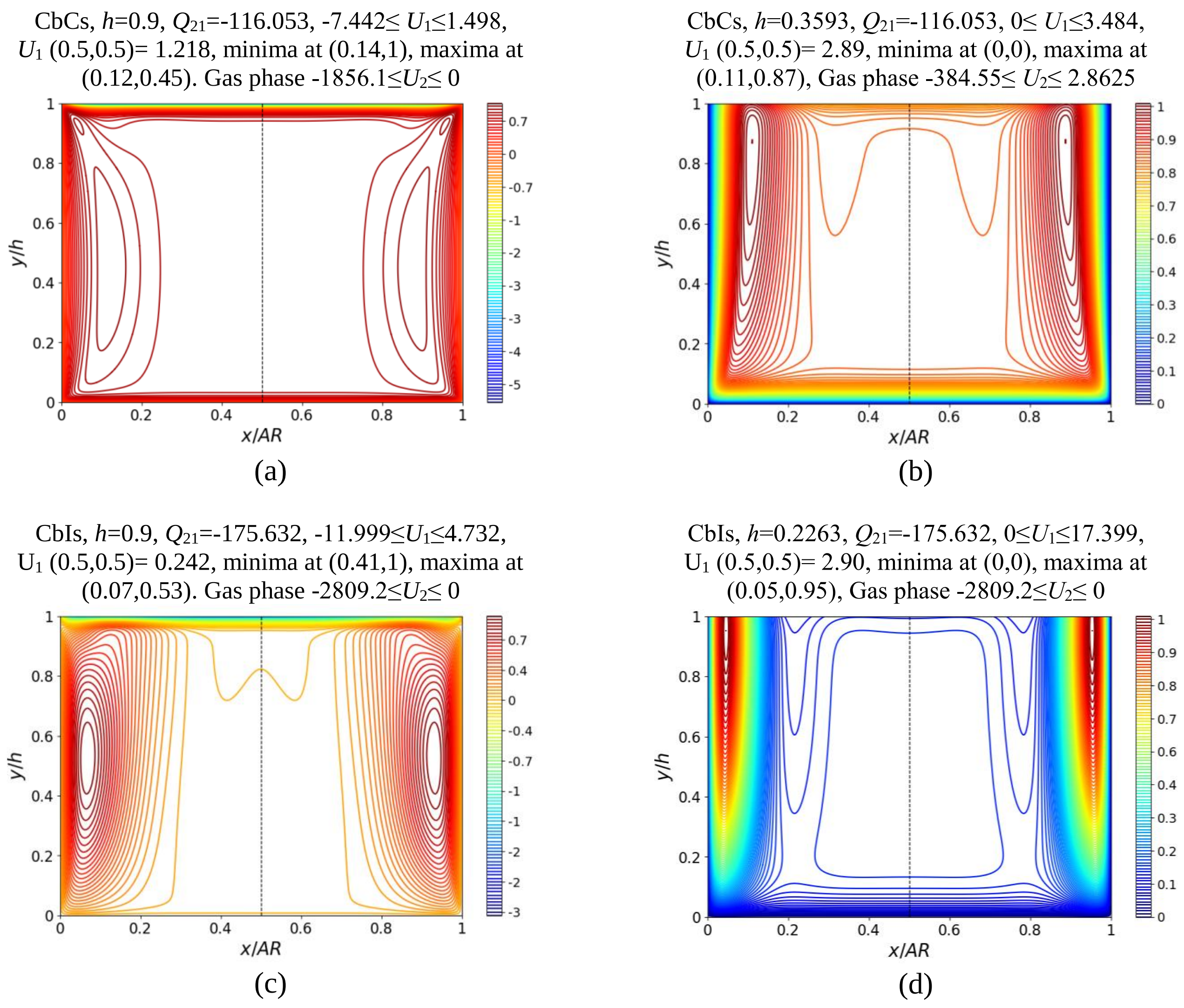

**Figure 14**: Velocity contours of the dual solutions in ducts with conducting bottom wall ($U_{1S}$ =0.1 *m/s*, *Ha*=103.625, $\beta$=5° and *AR*=1) (a-b) CbCs *h*=0.9 and *h*=0.36, (c-d) CbIs *h*=0.9 and *h*=0.23.

In the countercurrent flow region negative values of $\tilde{P}_f^0$, $P_f^1$ in Figures 17(a,b) indicate that the frictional pressure gradient is dominated by the upward gas flow, whereas positive values of these factors indicate that the frictional pressure gradient is dominated by the downward mercury flow. The latter occurs for the lower-holdup solution branch, which extends into the concurrent down flow region ($Q_{21}$ >0), where both $\tilde{P}_f^0$, $P_f^1$ remain positive. The downward flow of the mercury is obviously assisted by the hydrostatic pressure, $(dp/dz)_g > 0$ ($\tilde{P}_g^0$ <0, Figure 17c)., and when $dp/dz > 0$ , i.e., $-\tilde{P}_g^0 > \tilde{P}_f^0$, no external pump is needed to drive the air-mercury flow.

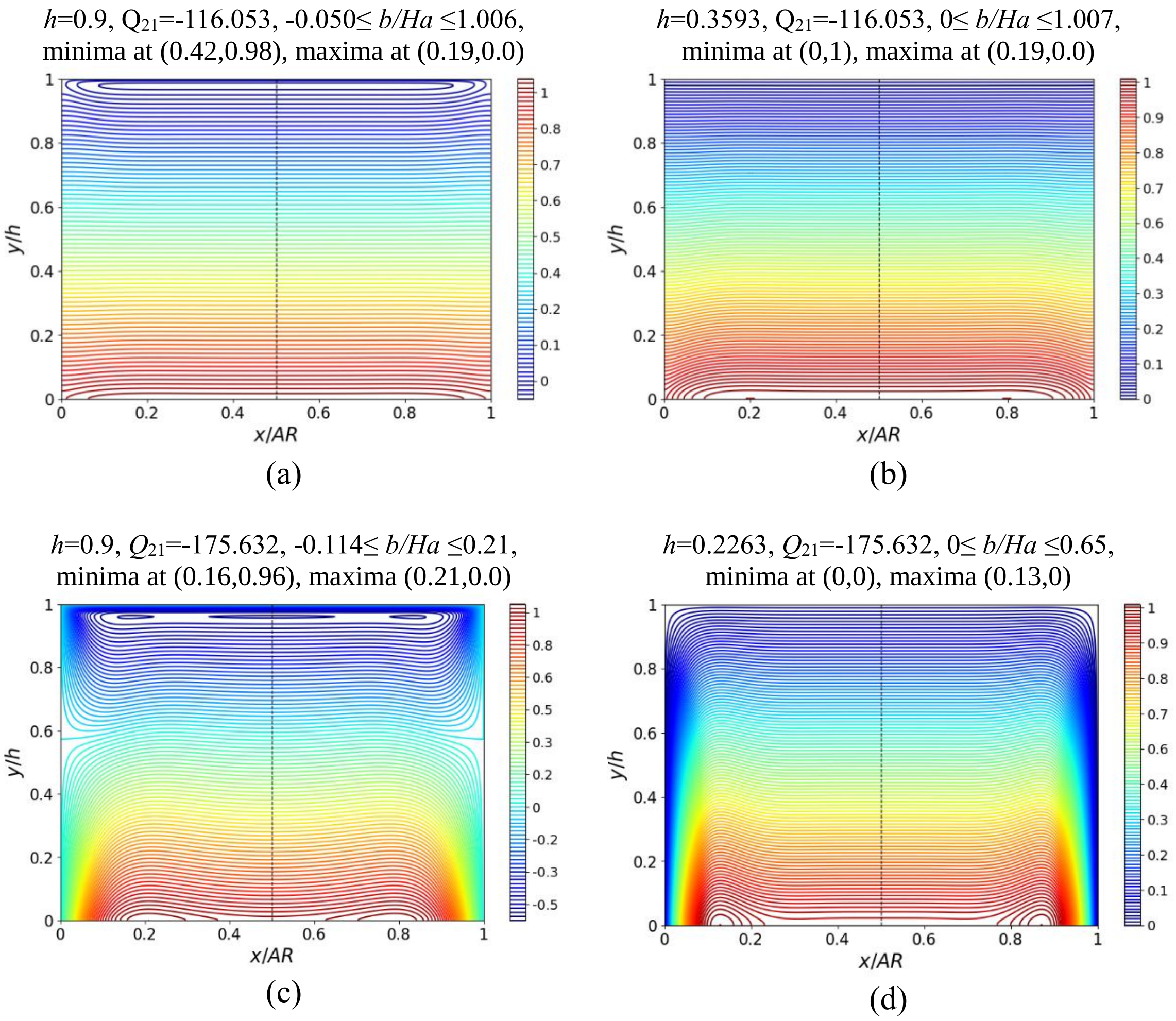

**Figure 15**: Induced magnetic field contours of the dual solutions in ducts with conducting bottom wall, associated with the velocity contours depicted in Figure 14 ($U_{1S}$ =0.1 *m/s*, $\beta$=5°, *Ha*=103.625, **AR**=1) (a-b) CbCs *h*=0.9 and *h*=0.36, (c-d) CbIs *h*=0.9 and *h*=0.23.

Inspection of Figure 17a shows that similarly to concurrent upward flow, the influence of the external magnetic field on the frictional pressure gradient is also the smallest in countercurrent flow and concurrent down flow for the fully insulating (IbIs) duct. As in the *Ha*=0 case, $\tilde{P}_f^0 < 0$ over a large portion of the countercurrent curve also for *Ha*=103.625 , indicating that the frictional pressure gradient is dominated by the upward flow of the gas phase. In the CbIs duct, only the upper holdup solution exhibits a frictional pressure gradient dominated by the gas phase, and the corresponding absolute values are lower than those obtained in IbIs duct for the same $Q_{21}$. Note that both IbIs and CbIs ducts a value of $\tilde{P}_f^0 = 0$ is obtained for a particular $Q_{21}$ (<0) along their lower solution branch, and values of $-1 < \tilde{P}_f^0 < 1$ are obtained in the vicinity of those points, indicating that the frictional pressure gradient in this ducts can be lower than obtained for single phase flow of mercury in the absence of magnetic field. The highest frictional pressure gradient is obtained in the fully conducting (CbCs) duct. In this case, the frictional pressure gradient is dominated by the downward mercury flow throughout the entire countercurrent and concurrent flow regions with $\tilde{P}_f^0$>1.

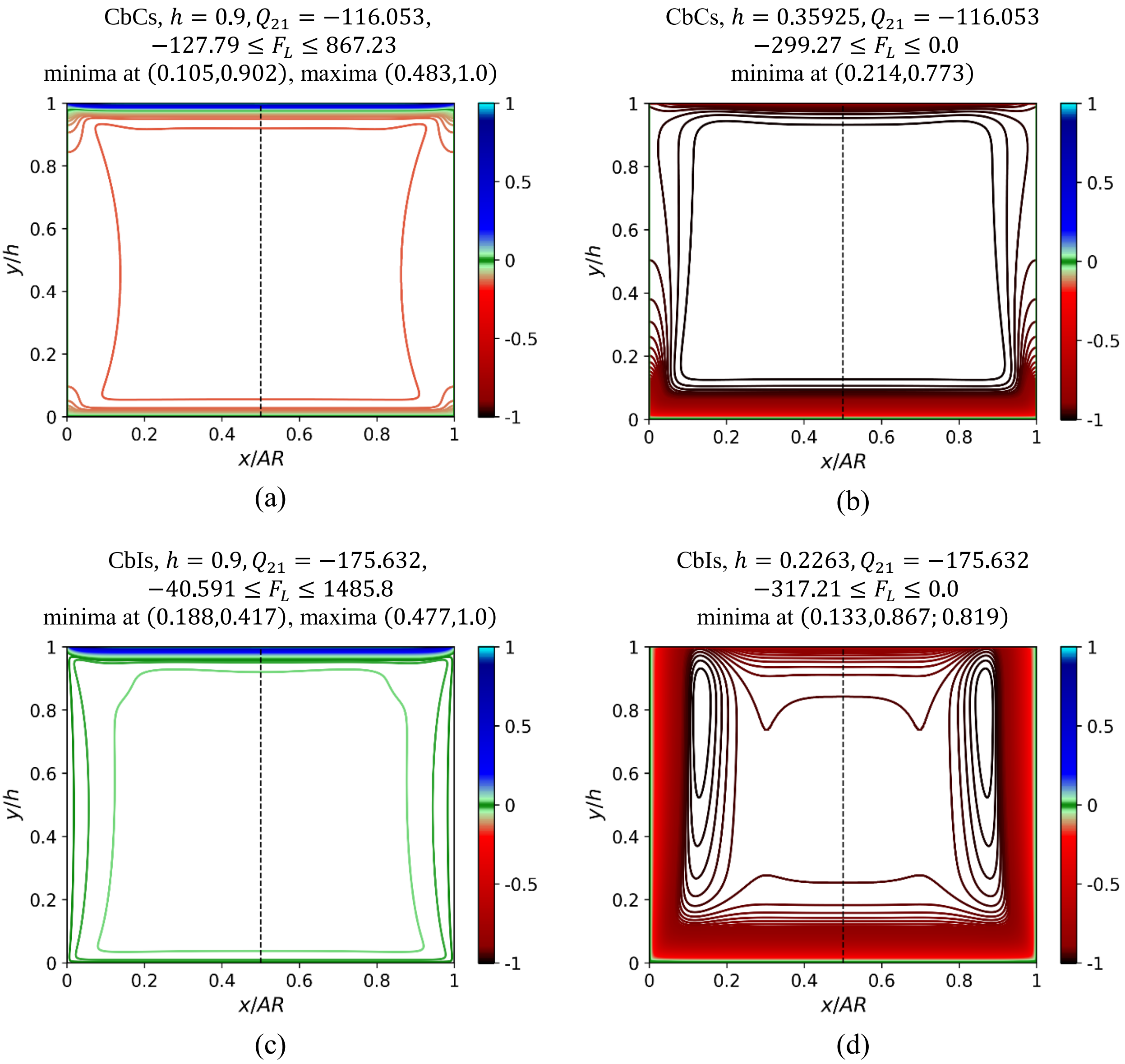

**Figure 16** The Lorentz force distribution of the dual solutions in ducts with conducting bottom wall corresponding to the velocity contours of the dual solutions depicted in Figure 14 Contours of the force normalized by its maximal absolute value, i.e., $F_L/\max(abs(F_L))$.

A reduction of the frictional pressure gradient compared to single phase flow of mercury under the same external magnetic field strength corresponds to $-1 < P_f^1 < 1$ in Figure 17b. Under strong magnetic field, the frictional pressure gradient in single phase mercury flow in ideally conducting CbCs ducts is substantially higher compared to that in a fully insulating duct ($\tilde{P}_f^0{}_{1s}$ =384.01 compared to 8.429, respectively). Consequently, introducing a counter current air flow produces a lubrication effect in the CbCs duct along both the lower and high holdup solution branches. A lubrication effect is obtained also in the IbIs and CbIs ducts over portions of their countercurrent flow region. Specifically, for the conditions examined in Figure 17b, lubrication effect is observed for a limited range of sufficiently low air flow rates along the upper solution branch, and in the in the vicinity of the $Q_{21}$ corresponding to $\tilde{P}_f^0, \tilde{P}_f^1 = 0$ along the lower solution branch. However, for all wall-conductivity configurations, $P_f^1 > 1$ for $Q_{21}$>0, indicating that introducing concurrent air flow increases the frictional pressure gradient in concurrent mercury-air downflow.

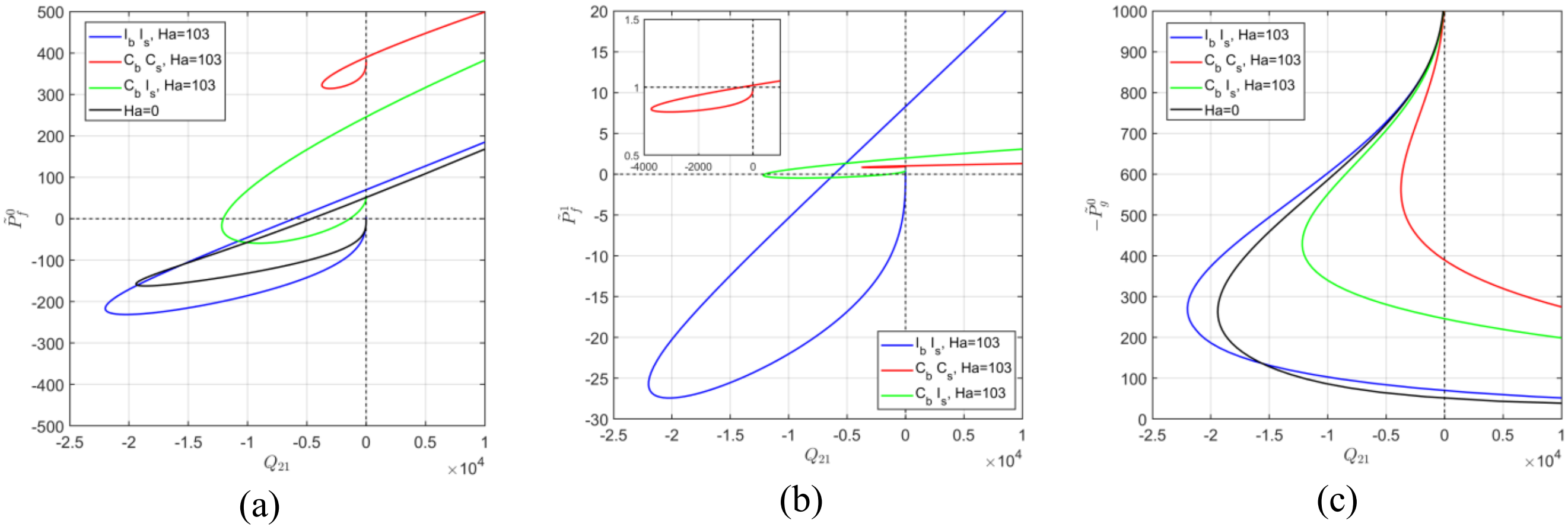

**Figure 17**: Variation of the pressure gradient with $Q_{21}$ in downward flow of mercury ($U_{1S}$ =0.1 *m/s*, *β*=5°, *Ha*=103.625 and *AR*=1). $Q_{21}$<0 corresponds to countercurrent flow (a) $\tilde{P}_f^0$, (b) $\tilde{P}_f^1 = \tilde{P}_f^0/\tilde{P}_f^0{}_{1s}$ (c)- $\tilde{P}_g^0$. The blue, red, green and black lines represent IbIs, CbCs, CbIs and *Ha*=0, respectively.

From the practical point of view, it is of interest to examine the effect of the air flow on total pressure gradient. Because the hydrostatic and frictional pressure gradients have opposite signs in single-phase flow of the conductive liquid, the interpretation of the normalized total pressure gradient in concurrent downward flow and in countercurrent flow is more complex than in concurrent upward flow ($\tilde{Y} < 0$ ). Therefore, for cases of $\tilde{Y} > 0$ , the total pressure gradient presented in Figure 18 is normalized by the hydrostatic pressure gradient of single-phase mercury flow, i.e., $\pi_T =$ $(\frac{dp}{dx})/(\rho_1 g sin\beta)$ . Positive values indicate that the specified mercury downward flowrate is entirely driven by gravity also in the presence of air upward gas flow and the specified liquid flow rate is maintained by a restriction at the liquid outlet.

Figures 18(a,b) show the effect of the *Ha* number and wall conductivity configuration on the $\pi_T$ values. The results demonstrate that even when the total pressure gradient is considered, the values obtained in insulating duct at high *Ha* remain similar to those obtained in the absence of magnetic field both in the concurrent down and countercurrent flow regions. On the other hand, when the bottom wall is conducting (CbIb and CbIs ducts) the *Ha* number significantly influences the total pressure gradient in both flow regions.

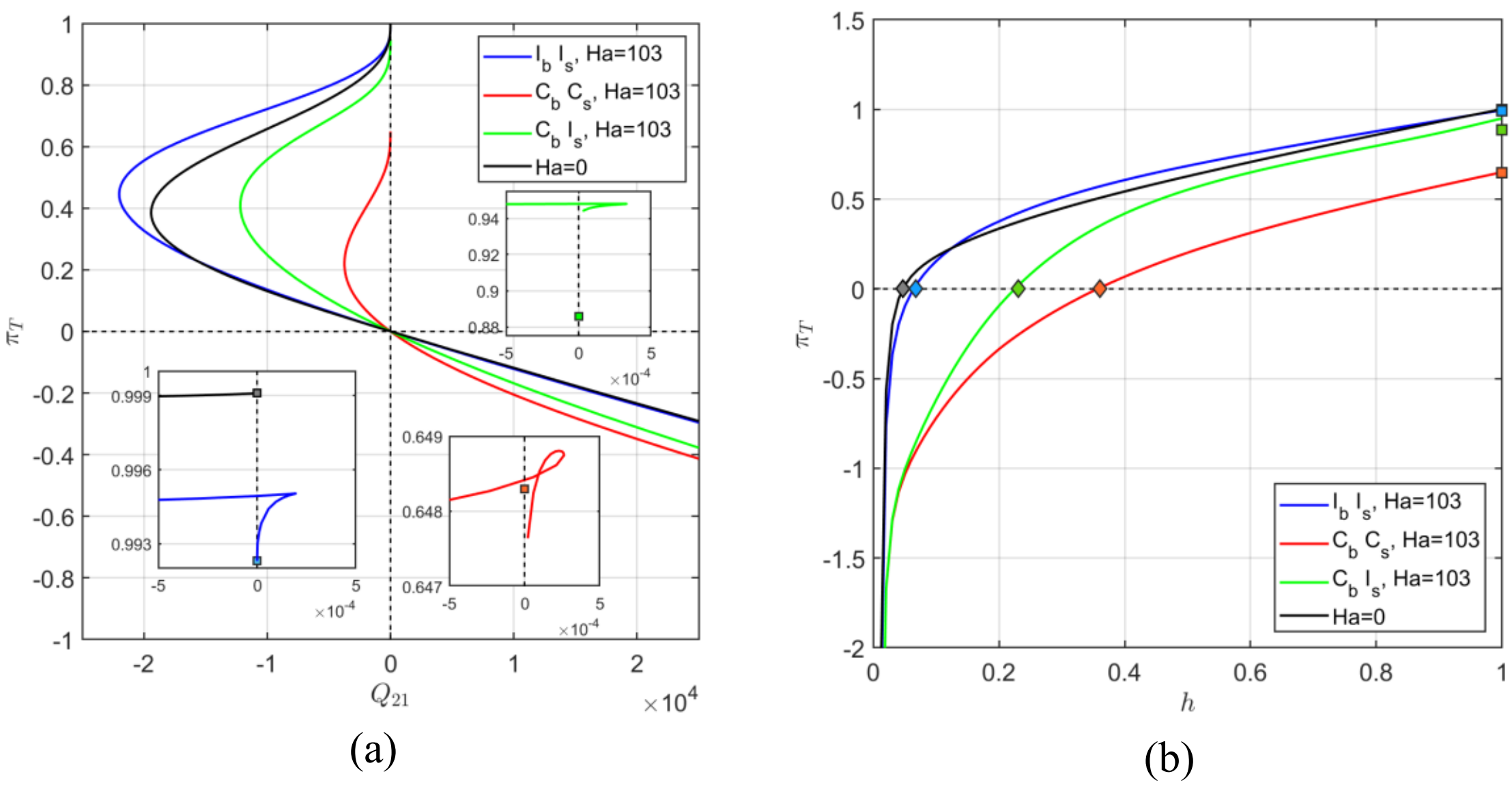

**Figure 18**: The total pressure gradient, $\pi_T$ in downward flow of mercury ($U_{1S}$ =0.1 *m/s*, *β*=5°, *Ha*=103.625 and *AR*=1). (a) Variation of $\pi_T$ with $Q_{21}$ (b) Variation of $\pi_T$ with the holdup. The blue, red, green and black lines represent IbIs, CbCs, CbIs and *Ha*=0, respectively.

For all wall conductivity configurations, positive values of $\pi_T$ are obtained in the counter current flow region, indicating that the specified mercury downward flowrate is entirely driven by gravity also in the presence of air upward gas flow. The holdup range corresponding to countercurrent flow extends from a value close to unity (see Figure 12b), to the holdup at which $\pi_T = 0$ (marked in Figure 18b by a square and diamond for the respected cases). The $\pi_T = 0$ is attained when the frictional pressure gradient is exactly balanced by the hydrostatic pressure gradient. For all wall conductivity configurations, $\pi_T = 0$ corresponds to the low holdup solution obtained for $Q_{21} = 0$ (see Figure 12b), where the air is circulating in the duct with zero net flow. This holdup is the highest for the CbCs duct, which is consistent with the narrower countercurrent region obtained for this duct. Negative $\pi_T$ values correspond to the concurrent down flow region. In fact, under conditions for which $\pi_T \leq 0$ for single phase liquid flow (i.e. holdup =1), a countercurrent flow region is infeasible.

The $\pi_T$ values corresponding to single phase mercury flow (*h*=1) are marked in Figure 18a by square in the inserted frames. These inserts enlarge the high-holdup region corresponding to the triple-holdup solutions in concurrent downward flow obtained for low $Q_{21} > 0$ For single phase flow $\pi_T \leq 1$ , since the direction frictional pressure gradient is opposite to the hydrostatic pressure gradient. Because the frictional pressure gradient in single-phase flow is the largest in the CbCs duct, the corresponding $\pi_T$ value is the lowest among the different wall conductivity configurations. It is worth noting that, as $h \to 1$, the gas–liquid values continuously approach the single-phase-flow value only in the IbIs duct, for which the boundary condition on the induced magnetic field is the same at the upper wall and at the gas–liquid interface.

In the concurrent downward flow region, the low holdup solution corresponds to $\pi_T \leq 0$ for of all wall conductivity configurations, indicating that a pump is required to drive the concurrent downward air- mercury flow. The required pumping power is, however, the lowest in the insulating duct. The two high holdup solutions in the triple-solution region of concurrent downward flow are associated with $\pi_T > 0$ . Thus, no pumping is required to drive the specified mercury flow in those high holdup configurations.

**Conclusions**

This study explores the two-phase flow characteristics of fully developed stratified MHD flow involving a conductive liquid and a non-conductive gas in inclined rectangular ducts subjected to external vertical magnetic field. The solutions for the liquid holdup, dimensionless pressure gradient, and velocity field, are governed by the Hartmann number (*Ha*), gas-to-liquid flow rate ratio ($Q_{21}$), viscosity ratio ($\eta_{21}$), inclination parameter $\tilde{Y}$, the duct aspect ratio (*AR*) and the conductivities of the bottom and side walls. In general, the governing problem requires numerical solution. Analytical solutions for the flow between two infinite plates (the TP model) considering either fully insulating or ideally conducting bottom wall. By neglecting the sidewalls, this model corresponds $AR \to \infty$. These analytical solutions provide valuable insights into the influence of the system parameters on the flow characteristics. However, because side-wall effects are neglected, the TP model predictions cannot be considered as quantitatively reliable for ducts of finite *AR*, particularly for *AR* of the order of unity. Analytical solutions, in the form of infinite series of hyperbolic functions, are also presented for rectangular ducts with insulating sidewalls and arbitrary bottom wall conductivity, which were used to validate the corresponding numerical solutions.

Mercury-air flow in a square duct has been selected as a case study to explore the MHD stratified flow characteristics in inclined ducts. The results demonstrate that gas–liquid stratified MHD flow in inclined rectangular ducts is governed by a strong interplay among gravity, Lorentz forces, as well as wall and interfacial shear stresses. Unlike single-phase MHD flow, duct inclination can profoundly alter two-phase flow because it changes the liquid holdup and therefore the relative importance of gravitational, frictional, and electromagnetic forces. The effect is particularly pronounced for liquid metals and gas flow because of the very large density difference between the phases.

A central finding is that wall conductivity is a key parameter controlling the influence of the magnetic field, which cannot be neglected even at very small magnetic Reynolds numbers. Because the gas phase is electrically nonconductive, the conductivity of the top wall has no effect. Under externally imposed vertical magnetic field, the system response is governed primarily by the conductivity of the bottom wall. When the bottom wall is insulating (IbIs duct), variations in the side-wall conductivity has only a minor effect, resulting in similar flow characteristics in IbIs and IbCs ducts remain similar. In contrast, when the bottom wall is conducting, the side-wall conductivity significantly affects the induced magnetic field, Lorentz-force distribution, velocity field, holdup, and pressure gradient. Consequently, CbCs and CbIs ducts can exhibit substantially different behavior despite both having a conducting bottom wall.

For concurrent upward flow, gravity strongly opposes the motion of the dense liquid and therefore produces high mercury holdups even at shallow inclination angles. Application of a magnetic field generally further increases the holdup by retarding the conductive liquid. Nevertheless, because gravity is dominant, the relative influence of the magnetic field and wall conductivity on holdup is weaker than in horizontal flow, particularly at moderate gas-to-liquid flow-rate ratios, $Q_{21}$. At sufficiently high $Q_{21}$, however, the holdup decreases sharply and a region with three possible holdup solutions may develop. The location and even the existence of this multiple-solution region depend strongly on the magnetic field and wall conductivity.

The velocity-field analysis shows that the magnetic field does considerably more than simply retard the liquid. Depending on the induced-field distribution, the Lorentz force can either oppose or assist the local liquid motion. This produces complex velocity structures, including local backflow and off-center velocity maxima. In upward flow, backflow can occur preferentially near the side walls rather than near the bottom wall, as predicted by the two-plate approximation. These backflow regions are potentially important because they can introduce disturbances near the duct inlet and may therefore promote instability of the stratified configuration.

In the triple-solution region of concurrent upward flow, the magnetic field substantially modifies the gravity-dominated high- and intermediate-holdup configurations. In the insulating duct, applying the magnetic field markedly reduces liquid backflow; where for the high-holdup solution, the maximum backflow velocity is reduced by approximately an order of magnitude. The low-holdup solution is dominated by gas-induced interfacial shear and is relatively insensitive to the magnetic field. Ducts with a conducting bottom wall do not exhibit triple solutions over the conditions examined because their stronger retarding Lorentz forces substantially modify the force balance.

The pressure-gradient analysis indicates that the frictional pressure gradient is the lowest in insulating duct and increases relatively modestly with *Ha* compared with ducts having a conducting bottom wall. However, the frictional-pressure-gradient response alone is insufficient for assessing the energetic benefit of gas injection in inclined flows. Although significant frictional lubrication can occur under some conditions, particularly in the CbIs configuration, the hydrostatic contribution remains important even at relatively shallow inclinations. When the total pressure gradient and the gas-flow pumping requirement are considered, introducing air generally does not reduce the overall pumping power in concurrent upward flow. Even the pronounced decrease in total pressure gradient associated with the sharp holdup reduction near the triple-solution region occurs only at very high gas flow rates. Consequently, a net pumping-power reduction through gas injection of a gas layer is impractical under the conditions investigated.

For countercurrent flow, the magnetic field generally reduces the range of upward gas flow rates for which steady countercurrent operation is possible because the Lorentz force weakens the relative influence of gravity on the liquid. This effect is again the weakest in the fully insulating duct: even at high *Ha*, its countercurrent-flow range can remain comparable to that without a magnetic field. On the other hand, a conducting bottom wall can suppress countercurrent flow completely unless the inclination parameter is sufficiently large. Replacing conducting side walls with insulating ones extends the feasible countercurrent-flow region, confirming the beneficial effect of electrically insulating side walls when the bottom wall is conducting.

A particularly important feature of countercurrent flow is the existence of two holdup solutions for the same phase flow rates. These branches merge at the flooding point, beyond which steady countercurrent flow is impossible. At sufficiently large inclination parameters, the upper branch can cross and extend into the concurrent downward-flow region, producing two additional high-holdup solutions. Thus, concurrent downward flow can also exhibit a narrow region with three possible holdups at very low gas flow rates.

The countercurrent velocity fields further demonstrate the strong effect of electromagnetic boundary conditions. With an insulating bottom wall, side-wall conductivity has little influence. With a conducting bottom wall, however, insulating the side walls can produce a highly non-uniform, jet-like liquid flow, in which most of the mercury is transported through high-velocity regions adjacent to the side walls while the core is nearly stagnant. This behavior results from strong electromagnetic braking in the liquid core combined with weaker Lorentz forces near the insulating side walls.

From an energetic perspective, countercurrent air flow can produce a frictional lubrication effect over parts of the countercurrent regime, including in the fully conducting duct because its single-phase MHD pressure gradient is particularly large. In contrast, concurrent downward air flow increases the frictional pressure gradient relative to single-phase mercury flow for all wall configurations considered.

Finally, the total pressure-gradient analysis clarifies the pumping requirements in countercurrent and concurrent down flow configurations. Positive pressure gradient throughout the countercurrent-flow region indicates that gravity alone can drive the prescribed downward mercury flow despite the opposing upward gas flow; flow regulation rather than pumping is therefore required on the liquid side. The condition of zero (dimensionless) total pressure gradient, $\pi_T = 0$, marks the balance between hydrostatic and frictional pressure gradients and coincides with the low-holdup solution at $Q_{21}$=0. In concurrent downward flow, the low-holdup branch corresponds to $\pi_T < 0$ , so pumping is required, although the required pumping power is lowest for the insulating duct. The two high-holdup solutions in the triple-solution region have $\pi_T > 0$ , hence the prescribed mercury flow can be maintained without pumping.

Overall, the study shows that the electromagnetic boundary conditions imposed by the duct walls fundamentally determine how the magnetic field modifies the balance among gravity, interfacial shear, viscous stresses, and Lorentz forces. Fully insulating ducts generally exhibit the weakest magnetic-field effects and behavior closest to the non-MHD gas-liquid flow, whereas conducting-bottom-wall configurations produce much stronger electromagnetic effects and greater sensitivity to side-wall conductivity. These interactions govern not only the liquid holdup and pressure gradient, but also the existence of multiple steady solutions, flooding limits, local backflow and jet-like structures, and ultimately the pumping requirements of the system.

**Author Contributions:** Conceptualization, A.G. and N.B.; Methodology I.B. and N.B. and A.G., Software, I.B., A.P. and A.G.; Validation, I.B., A.P., N.B and A.G.; Formal analysis, S.P., A.P.,

I.B., A.G. and N.B.; Resources, A.G. and N.B.; Data curation, I.B. S.P. and A.P.; Writing—original draft, N.B. and I.B; Writing—review & editing, A.P. and N.B.; Visualization, S.P. and I.B; Supervision, A.G and N.B.; Funding acquisition, A.G. and N.B. All authors have read and agreed to the published version of the manuscript.

**Funding:** This research was supported by Israel Science Foundation (ISF) grant No 1363/23 and by the Israel Ministry of Aliyah and Integration (for A. Parfenov).

**Data Availability Statement:** The original contributions presented in the study are included in the article, further inquiries can be directed to the corresponding author.

**Conflicts of Interest:** The authors declare no conflict of interest.

**Acknowledgments**

This research was supported by Israel Science Foundation (ISF) grant No 1363/23.

**Appendix A: Analytical Solutions**

Consider a problem with vertically directed magnetic field and fully insulating side walls. We follow similar steps as in Pal et al., (2026) with the exception of different $G_j$ for two phases instead of the same $G$. The dimensionless problem consists of the following equations:

$$\Delta u_2 = \eta_{12} G_2, \qquad (A.1a)$$

$$\Delta u_1 + Ha\frac{\partial b}{\partial y} = G_1, \qquad (A.1b)$$

$$\Delta b + Ha\frac{\partial u_1}{\partial y} = 0, \qquad (A.1c)$$

in the domain $(x, y) \in [0; AR] \times [0; 1]$. The conditions at the interface (located at $y = h$) are:

$$u_1 \underset{y=h}{=} u_2, \qquad \eta_{12}\frac{\partial u_1}{\partial y} \underset{y=h}{=} \frac{\partial u_2}{\partial y}, \qquad b \underset{y=h}{=} 0, \qquad (A.2)$$

and on the walls:

$$u_j(x = 0) = u_j(x = AR) = 0, \;\; b(x = 0) = b(x = AR) = 0, \qquad (A.3a)$$

$$u_1(y = 0) = u_2(y = 1) = 0, \qquad b - C_{bw}\frac{\partial b}{\partial y} \underset{y=0}{=} 0. \qquad (A.3b)$$

Using (A.3a), we can represent $u_j, b$ as:

$$u_j = \sum u_{j,k}(y)\sin(\lambda_k x), \qquad b = \sum b_k(y)\sin(\lambda_k x), \qquad \lambda_k = \pi k/AR, \qquad k = 1,2,\ldots \qquad (A.4)$$

Denote $f_{j,k} = -\gamma_k G_j$, where

$$1 \approx \sum_k \gamma_k \sin(\lambda_k x)\,, \tag{A.5a}$$

$$\gamma_k = \int_0^{AR} \sin(\lambda_k x)\,dx \Big/ \int_0^{AR} \sin^2(\lambda_k x)\,dx = \frac{2}{\pi k}[1-(-1)^k]. \tag{A.5b}$$

Then terms of the expansion (A.4) should satisfy:

$$u''_{2,k} - \lambda_k^2 u_{2,k} = -\eta_{12} f_{2,k}, \tag{A.6a}$$

$$u''_{1,k} - \lambda_k^2 u_{1,k} + Ha\, b'_k = -f_{1,k}, \tag{A.6b}$$

$$b''_k - \lambda_k^2 b_k + Ha\, u'_{1,k} = 0; \tag{A.6c}$$

with conditions on the walls and at the interface:

$$u_{1,k}(h) = u_{2,k}(h), \qquad \eta_{12} u'_{1,k}(h) = u'_{2,k}(h), \qquad b_k(h) = 0; \tag{A.7a}$$

$$u_{1,k}(0) = 0, \qquad u_{2,k}(1) = 0, \qquad b_k - C_{bw} b'_k \underset{y=0}{=} 0. \tag{A.7b}$$

While solving the problem for an individual term, let us omit index $k$. We denote

$$v_1 = u_1 + b, \qquad w_1 = u_1 - b, \tag{A.8}$$

so the problem becomes:

$$u''_2 - \lambda^2 u_2 = -\eta_{12} f_2, \tag{A.9a}$$

$$v''_1 - \lambda^2 v_1 + Ha\, v'_1 = -f_1, \tag{A.9b}$$

$$w''_1 - \lambda^2 w_1 - Ha\, w'_1 = -f_1; \tag{A.9c}$$

$$u_2(h) = v_1(h) = w_1(h), \qquad 2u'_2 \underset{y=h}{=} \eta_{12}(v'_1 + w'_1), \tag{A.10a}$$

$$u_2(1) = 0, \qquad v_1(0) + w_1(0) = 0, \qquad \left(1 - C_{bw}\frac{d}{dy}\right)(v_1 - w_1) \underset{y=0}{=} 0. \tag{A.10b}$$

In the case of horizontal flow (β=0) G1=G2=G and $f_1 = f_2 = f$.

Now denote

$$M = Ha/2, \qquad N = \sqrt{\lambda^2 + M^2}, \tag{A.11}$$

so that $Ha = 2M,\ \lambda^2 = N^2 - M^2$, and equations (A.9a-c) can be written as

$$\left(\frac{d}{dy} + \lambda\right)\left(\frac{d}{dy} - \lambda\right) u_2 = -\eta_{12} f_2, \tag{A.12a}$$

$$\left(\frac{d}{dy} + M + N\right)\left(\frac{d}{dy} + M - N\right) v_1 = -f_1, \tag{A.12b}$$

$$\left(\frac{d}{dy} - M + N\right)\left(\frac{d}{dy} - M - N\right) w_1 = -f_1. \tag{A.12c}$$

Given that, and considering no-slip conditions (A.10b), the general solution for $u_2, v_1, w_1$ can be written:

$$u_2 = \frac{\eta_{12}}{\lambda^2}\left(f_2[1 - \cosh(\lambda[1-y])] + C^u \sinh(\lambda[1-y])\right), \qquad (A.13a)$$

$$v_1 = \frac{1}{\lambda^2}\left(f_1 - (f_1 - D)\, e^{-My}\cosh(Ny) + C^v e^{-My}\sinh(Ny)\right), \qquad (A.13b)$$

$$w_1 = \frac{1}{\lambda^2}\left(f_1 - (f_1 + D)\, e^{+My}\cosh(Ny) + C^w e^{+My}\sinh(Ny)\right), \qquad (A.13c)$$

where $D, C^u, C^v, C^w$ are unknown constants. The condition for $v_1 - w_1$ on the lower wall results in:

$$D = C_{bw}\left[ Mf_1 + N\frac{C^v - C^w}{2}\right]. \qquad (A.14)$$

Using (A.10a), we introduce constant $A$ such that $u_2(h) = v_1(h) = w_1(h) = A/\lambda^2$, then:

$$-f_2\cosh(\lambda[1-h]) + C^u\sinh(\lambda[1-h]) = A/\eta_{12} - f_2, \qquad (A.15a)$$

$$-(f_1 - D)\ \cosh(Nh) + C^v\sinh(Nh) = (A - f_1)e^{+Mh}, \qquad (A.15b)$$

$$-(f_1 + D)\ \cosh(Nh) + C^w\sinh(Nh) = (A - f_1)e^{-Mh}. \qquad (A.15c)$$

From (A.15b, A.15c) we get:

$$\frac{C^v + C^w}{2}\cdot\sinh(Nh) = (A - f_1)\cosh(Mh) + f_1\cosh(Nh), \qquad (A.16a)$$

$$D\cosh(Nh) + \frac{C^v - C^w}{2}\cdot\sinh(Nh) = (A - f_1)\sinh(Mh). \qquad (A.16b)$$

Substituting (A.14) into (A.16b):

$$\frac{C^v - C^w}{2} = (A - f_1)\cdot\frac{\sinh(Mh)}{\sinh(Nh)}\cdot\frac{\tanh(Nh)}{\tanh(Nh) + C_{bw}N} - f_1\cdot\frac{C_{bw}M}{C_{bw}N + \tanh(Nh)}. \qquad (A.17)$$

Let us denote:

$$X = \frac{C_{bw}}{C_{bw} + \frac{1}{N}\tanh(Nh)}; \qquad (A.18)$$

Then (A.17) is rewritten as:

$$\frac{C^v - C^w}{2} = (A - f_1)\cdot\frac{\sinh(Mh)}{\sinh(Nh)}\cdot(1 - X) - f_1\cdot\frac{M}{N}\cdot X. \qquad (A.19)$$

Note that $X = 0$ for perfectly insulating bottom wall, $X = 1$ for perfectly conductive wall, and depends on $N$ otherwise. Constants $D, C^u, C^v, C^w$ can be represented using $A$:

$$C^u = \frac{A/\eta_{12} - f_2}{\sinh(\lambda[1-h])} + f_2\frac{\cosh(\lambda[1-h])}{\sinh(\lambda[1-h])}, \qquad (A.20a)$$

$$C^v = (A - f_1)\frac{e^{+Mh}}{\sinh(Nh)} + f_1\frac{\cosh(Nh)}{\sinh(Nh)} + X\left[f_1\left(\frac{\sinh(Mh)}{\sinh(Nh)} - \frac{M}{N}\right) - A\,\frac{\sinh(Mh)}{\sinh(Nh)}\right], \qquad (A.20b)$$

$$C^w = (A - f_1)\frac{e^{-Mh}}{\sinh(Nh)} + f_1\frac{\cosh(Nh)}{\sinh(Nh)} - X\left[f_1\left(\frac{\sinh(Mh)}{\sinh(Nh)} - \frac{M}{N}\right) - A\ \frac{\sinh(Mh)}{\sinh(Nh)}\right], \qquad (A.20c)$$

$$D = X \cdot \tanh(Nh)\left[A\ \frac{\sinh(Mh)}{\sinh(Nh)} - f_1\left(\frac{\sinh(Mh)}{\sinh(Nh)} - \frac{M}{N}\right)\right], \qquad (A.20d)$$

which can be found from shear stress condition $2u_2' \underset{y=h}{=} \eta_{12}(v_1' + w_1')$:

$$\begin{aligned} &f_2\lambda\sinh(\lambda[1-h]) - \lambda C^u\cosh(\lambda[1-h]) = \\ &= -N(f_1\cosh(Mh) + D\sinh(Mh))\sinh(Nh) + N\frac{C^v e^{-Mh} + C^w\ e^{+Mh}}{2}\cosh(Nh). \end{aligned} \qquad (A.21)$$

Substituting (A.20a-d) into (A.21) yields expression for $A$:

$$A = \frac{f_2\lambda\ \dfrac{\cosh(\lambda[1-h]) - 1}{\sinh(\lambda[1-h])} + f_1 N\dfrac{\cosh(Nh) - \cosh(Mh)}{\sinh(Nh)} + Xf_1N\dfrac{\sinh(Mh)}{\cosh(Nh)}\left(\dfrac{\sinh(Mh)}{\sinh(Nh)} - \dfrac{M}{N}\right)}{\dfrac{\lambda}{\eta_{12}}\dfrac{\cosh(\lambda[1-h])}{\sinh(\lambda[1-h])} + N\dfrac{\cosh(Nh)}{\sinh(Nh)} + XN\dfrac{\sinh(Mh)}{\cosh(Nh)}\cdot\dfrac{\sinh(Mh)}{\sinh(Nh)}}. \qquad (A.22)$$

We also substitute (A.20a-d) into (A.13a-c) and find $u_1 = (v_1 + w_1)/2$, $b_1 = (v_1 - w_1)/2$:

$$u_2 = \frac{f_2\eta_{12}}{\lambda^2}\left(1 - \frac{\sinh(\lambda[y-h])}{\sinh(\lambda[1-h])}\right) + \frac{1}{\lambda^2}(A - f_2\eta_{12})\ \frac{\sinh(\lambda[1-y])}{\sinh(\lambda[1-h])}, \qquad (A.23a)$$

$$\begin{aligned} u_1 = &\frac{f_1}{\lambda^2}\left(1 - \cosh(My)\cdot\frac{\sinh(N[h-y])}{\sinh(Nh)}\right) + \frac{1}{\lambda^2}(A - f_1)\cosh(M[h-y])\cdot\frac{\sinh(Ny)}{\sinh(Nh)} - \\ &-\frac{X}{\lambda^2}\left(f_1\cdot\frac{M}{N} + (A - f_1)\cdot\frac{\sinh(Mh)}{\sinh(Nh)}\right)\sinh(My)\cdot\frac{\sinh(N[h-y])}{\cosh(Nh)}, \end{aligned} \qquad (A.23b)$$

$$\begin{aligned} b_1 = &\frac{f_1}{\lambda^2}\ \sinh(My)\cdot\frac{\sinh(N[h-y])}{\sinh(Nh)} + \frac{1}{\lambda^2}(A - f_1)\sinh(M[h-y])\cdot\frac{\sinh(Ny)}{\sinh(Nh)} + \\ &+\frac{X}{\lambda^2}\left(f_1\cdot\frac{M}{N} + (A - f_1)\cdot\frac{\sinh(Mh)}{\sinh(Nh)}\right)\cosh(My)\cdot\frac{\sinh(N[h-y])}{\cosh(Nh)}. \end{aligned} \qquad (A.23c)$$

Note that in deriving expressions (A.23b-c), we used identity $\cosh^2(Nh) - \sinh^2(Nh) = 1$. For large $N$ evaluating this identity can lead to numerical difficulties since finite precision arithmetic can cause this difference to evaluate to 0 instead of 1. Therefore, formulas (A.23a-A.23c) are preferable for numerical computations over the corresponding solution presented in Pal et al. (2026) for the case of horizontal ducts. Furthermore, if overflow should be avoided, these expressions can be rewritten without positive exponents as follows:

$$X = \frac{C_{bw}}{C_{bw} + \dfrac{1}{N}\cdot\dfrac{1 - e^{-2Nh}}{1 + e^{-2Nh}}} \qquad (A.24a)$$

$$A=\frac{\begin{bmatrix} f_1\left[\frac{\left(1+e^{-2Nh}\right)}{\left(1-e^{-2Nh}\right)}-e^{-(N-M)h}\frac{\left(1+e^{-2Mh}\right)}{\left(1-e^{-2Nh}\right)}\right]+\frac{f_2\lambda}{N}\left[\frac{\left(1+\mathrm{e}^{-2\lambda[1-h]}\right)}{\left(1-\mathrm{e}^{-2\lambda[1-h]}\right)}-\frac{2\,e^{-\lambda[1-h]}}{\left(1-\mathrm{e}^{-2\lambda[1-h]}\right)}\right]+ \\ +X\,f_1 e^{-(N-M)h}\cdot\frac{1-e^{-2Mh}}{1+e^{-2Nh}}\cdot\left(e^{-(N-M)h}\cdot\frac{1-e^{-2Mh}}{1-e^{-2Nh}}-\frac{M}{N}\right)\end{bmatrix}}{\frac{\left(1+e^{-2Nh}\right)}{\left(1-e^{-2Nh}\right)}+\frac{\lambda}{\eta_{12}N}\cdot\frac{\left(1+\mathrm{e}^{-2\lambda[1-h]}\right)}{\left(1-\mathrm{e}^{-2\lambda[1-h]}\right)}+X\,e^{-2(N-M)h}\cdot\frac{\left(1-e^{-2Mh}\right)^2}{1-e^{-4Nh}}}\qquad(A.24b)$$

$$u_2=\frac{f_2\eta_{12}}{\lambda^2}\left(1-e^{-\lambda(1-y)}\cdot\frac{1-e^{-2\lambda(y-h)}}{1-e^{-2\lambda(1-h)}}\right)+\frac{1}{\lambda^2}(A-f_2\eta_{12})e^{-\lambda(y-h)}\cdot\frac{1-e^{-2\lambda(1-y)}}{1-e^{-2\lambda(1-h)}}\qquad(A.25a)$$

$$\begin{aligned} u_1=&\frac{f_1}{\lambda^2}\left(1-\frac{e^{-(N-M)y}}{2}\cdot\frac{\left(1+e^{-2My}\right)\cdot\left(1-e^{-2N(h-y)}\right)}{1-e^{-2Nh}}\right)+\\ &+\frac{1}{\lambda^2}(A-f_1)\;\frac{e^{-(N-M)(h-y)}}{2}\cdot\frac{\left(1+e^{-2M(h-y)}\right)\cdot\left(1-e^{-2Ny}\right)}{1-e^{-2Nh}}-\\ &-\frac{X}{\lambda^2}\left(f_1\cdot\frac{M}{N}+(A-f_1)\cdot e^{-(N-M)h}\frac{1-e^{-2Mh}}{1-e^{-2Nh}}\right)\frac{e^{-(N-M)y}}{2}\cdot\frac{\left(1-e^{-2My}\right)\left(1-e^{-2N(h-y)}\right)}{1-e^{-2Nh}}\end{aligned}\qquad(A.25b)$$

$$\begin{aligned} b_1=&\frac{f_1}{\lambda^2}\frac{e^{-(N-M)y}}{2}\cdot\frac{\left(1-e^{-2My}\right)\cdot\left(1-e^{-2N(h-y)}\right)}{1-e^{-2Nh}}+\\ &+\frac{1}{\lambda^2}(A-f_1)\;\frac{e^{-(N-M)(h-y)}}{2}\cdot\frac{\left(1-e^{-2M(h-y)}\right)\cdot\left(1-e^{-2Ny}\right)}{1-e^{-2Nh}}\\ &+\frac{X}{\lambda^2}\left(f_1\cdot\frac{M}{N}+(A-f_1)\cdot e^{-(N-M)h}\frac{1-e^{-2Mh}}{1-e^{-2Nh}}\right)\frac{e^{-(N-M)y}}{2}\cdot\frac{\left(1+e^{-2My}\right)\left(1-e^{-2N(h-y)}\right)}{1-e^{-2Nh}}\end{aligned}\qquad(A.25c)$$

The flow rates $Q_1, Q_2$ are obtained by integrating phases' velocities over corresponding cross-sections:

$$Q_1=\int_0^{AR}\int_0^h u_1(x,y)\,dx\,dy=\int_0^{AR}\int_0^h\sum_k u_{1,k}(y)\sin(\lambda_k x)\;dx\,dy=\sum_k\frac{[1-(-1)^k]}{\lambda_k}\;q_{1,k}\,,\qquad(A.26a)$$

$$Q_2=\int_0^{AR}\int_h^1 u_2(x,y)\,dx\,dy=\int_0^{AR}\int_h^1\sum_k u_{2,k}(y)\sin(\lambda_k x)\;dx\,dy=\sum_k\frac{[1-(-1)^k]}{\lambda_k}\;q_{2,k}\,,\qquad(A.26b)$$

where

$$\begin{aligned} q_{1,k}=&\int_0^h u_{1,k}(y)dy=\frac{f_{1,k}h}{\lambda_k^2}+\frac{N_k\left(A_k-2f_{1,k}\right)}{\lambda_k^4}\cdot\left[\tanh\left(\frac{N_k h}{2}\right)-\frac{\cosh(Mh)-1}{\sinh(N_k h)}\right]+\\ &+X_k\cdot\frac{N_k A_k}{\lambda_k^4}\left(\frac{\sinh(Mh)}{\sinh(N_k h)}-\frac{M}{N_k}\right)\cdot\frac{\sinh(Mh)}{\cosh(N_k h)}-X_k\cdot\frac{N_k f_{1,k}}{\lambda_k^4}\left(\frac{\sinh(Mh)}{\sinh(N_k h)}-\frac{M}{N_k}\right)^2\cdot\tanh(N_k h)\,,\end{aligned}\qquad(A.27a)$$

$$q_{2,k}=\int_h^1 u_{2,k}(y)dy=\frac{f_{2,k}\eta_{12}(1-h)}{\lambda_k^2}+\frac{A_k-2f_{2,k}\eta_{12}}{\lambda_k^3}\cdot\tanh\left(\frac{\lambda_k}{2}[1-h]\right).\qquad(A.27b)$$

Or, in a different form:

$$q_{1.k}=\frac{f_{1,k}h}{\lambda_k^2}+\frac{N_k(A_k-2f_{1.k})}{\lambda_k^4}\cdot\left[\frac{1-e^{-N_kh}}{1+e^{-N_kh}}-\frac{e^{-(N_k-M)h}-2e^{-N_kh}+e^{-(N_k+M)h}}{1-e^{-2N_kh}}\right]+$$
$$+X_k\cdot\frac{N_kA_k}{\lambda_k^4}\left(e^{-(N_k-M)h}\cdot\frac{1-e^{-2Mh}}{1-e^{-2N_kh}}-\frac{M}{N_k}\right)\cdot e^{-(N_k-M)h}\cdot\frac{1-e^{-2Mh}}{1+e^{-2N_kh}}- \qquad (A.28a)$$
$$-X_k\cdot\frac{N_kf_{1,k}}{\lambda_k^4}\left(e^{-(N_k-M)h}\cdot\frac{1-e^{-2Mh}}{1-e^{-2N_kh}}-\frac{M}{N_k}\right)^2\cdot\frac{1-e^{-2N_kh}}{1+e^{-2N_kh}},$$

$$q_{2,k}=\frac{f_{2,k}\eta_{12}(1-h)}{\lambda_k^2}+\frac{A_k-2f_{2,k}\eta_{12}}{\lambda_k^3}\cdot\frac{1-e^{-\lambda_k(1-h)}}{1+e^{-\lambda_k(1-h)}}. \qquad (A.28b)$$

Note that for large enough , $A=O(f_1+f_2)$, $f_j=O(1/\lambda_k)$ and $\lambda_k=\frac{\pi k}{AR}=O(k)$, so that (A.26a-b) converge with terms of order $O(1/k^4)$. Also note that $(A.28a)$ involves multipliers $1/(1-e^{-2N_kh})$, so $N_kh$ should also be large enough. Hence, more terms could be required for smaller $h$.